\documentclass[preprint,12pt]{elsarticle}

\usepackage{amsmath,amsfonts,amssymb,bm}
\usepackage{epsfig,empheq,mathrsfs,mathtools,stmaryrd}
\usepackage{caption}
\usepackage{lineno} 
\usepackage{graphicx,subfigure}
\usepackage[title,titletoc,header]{appendix}
\usepackage{color}
\usepackage{import}
\usepackage{tikz}
\usepackage{siunitx}
\usepackage{graphicx}
\usepackage{dcolumn}
\usepackage{bm}

   \usepackage[utf8]{inputenc}
    \usepackage[]{empheq}
    \usepackage{cases}

\newcommand{\virgolette}[1]{``#1''}
 
\usepackage{comment}

 \usepackage{scalerel,stackengine,wasysym}
\stackMath
\newcommand\reallywidehat[1]{%
\savestack{\tmpbox}{\stretchto{%
  \scaleto{%
    \scalerel*[\widthof{\ensuremath{#1}}]{\kern-.6pt\bigwedge\kern-.6pt}%
    {\rule[-\textheight/2]{1ex}{\textheight}}
  }{\textheight}%
}{0.5ex}}%
\stackon[1pt]{#1}{\tmpbox}%
}

\usepackage{geometry}
\biboptions{numbers,sort&compress}

\usepackage[multiple]{footmisc}
\usepackage{bigfoot}

\DeclareNewFootnote{AAffil}[arabic]
\DeclareNewFootnote{ANote}[fnsymbol]

\usepackage{etoolbox}
\patchcmd{\emailauthor}{(#2)}{}{}{}
\patchcmd{\urlauthor}{(#2)}{}{}{}

\usepackage{changepage}
\usepackage{multirow}

\newlength{\tempheight}

\journal{Arxiv}

\usepackage{chapterbib}

\begin{document}

\begin{frontmatter}

\title{Sensitivity of viscoelastic characterization in multi-harmonic atomic force microscopy}

\author[add1]{Abhilash Chandrashekar\footnoteAAffil{These two authors contributed equally}}
\author[add1]{Arthur Givois\corref{cor1}$^{1}$}
\cortext[cor1]{Corresponding authors.}
\ead{arthur.givois@junia.com}
\author[add2]{Pierpaolo Belardinelli}
		\author[add1]{Casper L. Penning}
	\author[add1]{Alejandro M. Arag\'{o}n}
	\author[add1]{Urs Staufer}
	\author[add1]{Farbod Alijani\corref{cor1}}
\ead{f.alijani@tudelft.nl}
\address[add1]{Precision and Microsystems Engineering, TU Delft, Delft, The Netherlands}
\address[add2]{DICEA, Polytechnic University of Marche, Ancona, Italy}

\begin{abstract}
Quantifying the nanomechanical properties of soft-matter using multi-frequency atomic force microscopy (AFM) is crucial for studying the performance of polymers, ultra-thin coatings, and biological systems. Such  characterization processes often make use of cantilever's spectral components to discern nanomechanical properties within a multi-parameter optimization problem. This could inadvertently lead to an over-determined parameter estimation with no clear relation between the identified parameters and their influence on the experimental data. In this work, we explore the sensitivity of viscoelastic characterization in polymeric samples to the experimental observables of  multi-frequency intermodulation AFM. By performing simulations and experiments we show that surface viscoelasticity has negligible effect on the experimental data and can lead to inconsistent and often non-physical identified parameters. Our analysis reveals that this lack of influence of the surface parameters relates to a vanishing gradient and non-convexity while minimizing the objective function. By removing the surface dependency from the model, we show that the characterization of bulk properties can be achieved with ease and without any ambiguity.  Our work sheds light on the sensitivity issues that can be faced when optimizing for a large number of parameters and observables in AFM operation, and calls for the development of new viscoelastic models at the nanoscale and improved computational methodologies for nanoscale mapping of viscoelasticity using AFM. 
\end{abstract}

%
%

\end{frontmatter}

\section{Introduction}


Viscoelastic characterization  of soft-matter at the nanoscale is important for understanding cell membrane functioning \cite{raman2011mapping,Uhlig2017,efremov2017measuring,efremov2020measuring}, developing innovative materials in polymer science \cite{Chiefari1998,Chyasnavichyus2014,proksch2016practical}, and for advancing nanolithography \cite{Garcia2014,Wang2018}. In this regard, dynamic atomic force microscopy (AFM) has emerged as an indispensable tool for characterizing nanomechanical properties of soft matter, offering diverse operating conditions under which a wide variety of samples can be probed with gentle forces \cite{garcia2020nanomechanical,collinson2021best}.

Dynamic AFM imaging offers multiple observable channels in the form of higher harmonics, modal amplitude, and phase contrast signals to map nanomechanical properties. Among multi-harmonic AFM techniques, the emergence of  bi-modal and  intermodulation AFM (IM-AFM)  has led to a drastic increase in the number of experimental observables and a consequent advancement in our understanding of material properties at the nanoscale. In particular, IM-AFM extends the concept of multi-frequency observables by providing a fast and convenient method to measure a set of frequency components in a narrow frequency band centered around the fundamental resonance of the AFM cantilever \cite{platz2008intermodulation,platz2013interpreting}. These frequency components directly benefit from the mechanical resonance gain of the first mode and can be easily converted to tip-sample force quadratures, which are in turn linked to the conservative and dissipative interactions with a sample \cite{platz2013interpreting,platz2013interaction}. 

Despite the advancements in AFM instrumentation and the abundance of viscoelastic models at hand \cite{Ting1966,attard2001interaction,rajabifar2018dynamic,solares2014probing,thoren2018modeling, attard2001interaction,Rajabifar_2021}, a consistent and robust estimation of viscoelasticity using AFM has remained a challenge \cite{efremov2020measuring}. This is mainly due to the fact that the compositional contrast of AFM images depend on several nanomechanical properties including elasticity, surface relaxation, and adhesion. Untangling these effects from one another requires setting up an optimization problem, where a large parameter space has to be searched to minimize the error between the simulations from a model and experimental data. But, similar to any optimization problem, the insensitivity of the model parameters with respect to the measurement data on one side, and the non-convexity of the objective function on the other side, can lead to non-unique and often non-physical estimation of parameters. Therefore, knowledge about the sensitivity of the model parameters to AFM observable channels is of paramount importance to extract consistent and reliable viscoelastic properties in dynamic AFM applications. 

In this article we discuss the sensitivity issues that can arise when characterizing viscoelasticity using multi-frequency IM-AFM. We perform measurements on a polymer blend made of stiff Polystyrene (PS) and soft Low-Density-Polyethylene (LDPE), and use a moving surface model \cite{haviland2015probing,thoren2018modeling} to extract the bulk and the surface viscoelasticity. The estimation of viscoelastic properties is achieved by matching the experimental spectral components of tip-sample force to the ones predicted by a computational model via an optimization procedure. To ascertain the sensitivity of the model parameters on the physical observables,  we perform a comprehensive comparison involving both local and global optimization techniques, and reveal a lack of sensitivity of surface motion to the experimental data obtained from IM-AFM. We show that the issue of insensitivity manifests itself during the optimization of the objective function by means of a vanishing gradient with respect to the surface parameters. To overcome this problem, we introduce a simple model, neglecting surface motion, which leads to statistically consistent and robust identification of bulk viscoelastic parameters.This work thus provides a general framework that can be used for investigating the reliability of similar viscoelastic models used for nanomechanical characterization in multi-frequency AFM applications.

\section{Experimental Results}\label{sec:Expe}
\begin{figure}[htbp]
	\sbox0{
		\begin{minipage}{\columnwidth}
			\centering
			\includegraphics[width=0.66\columnwidth]{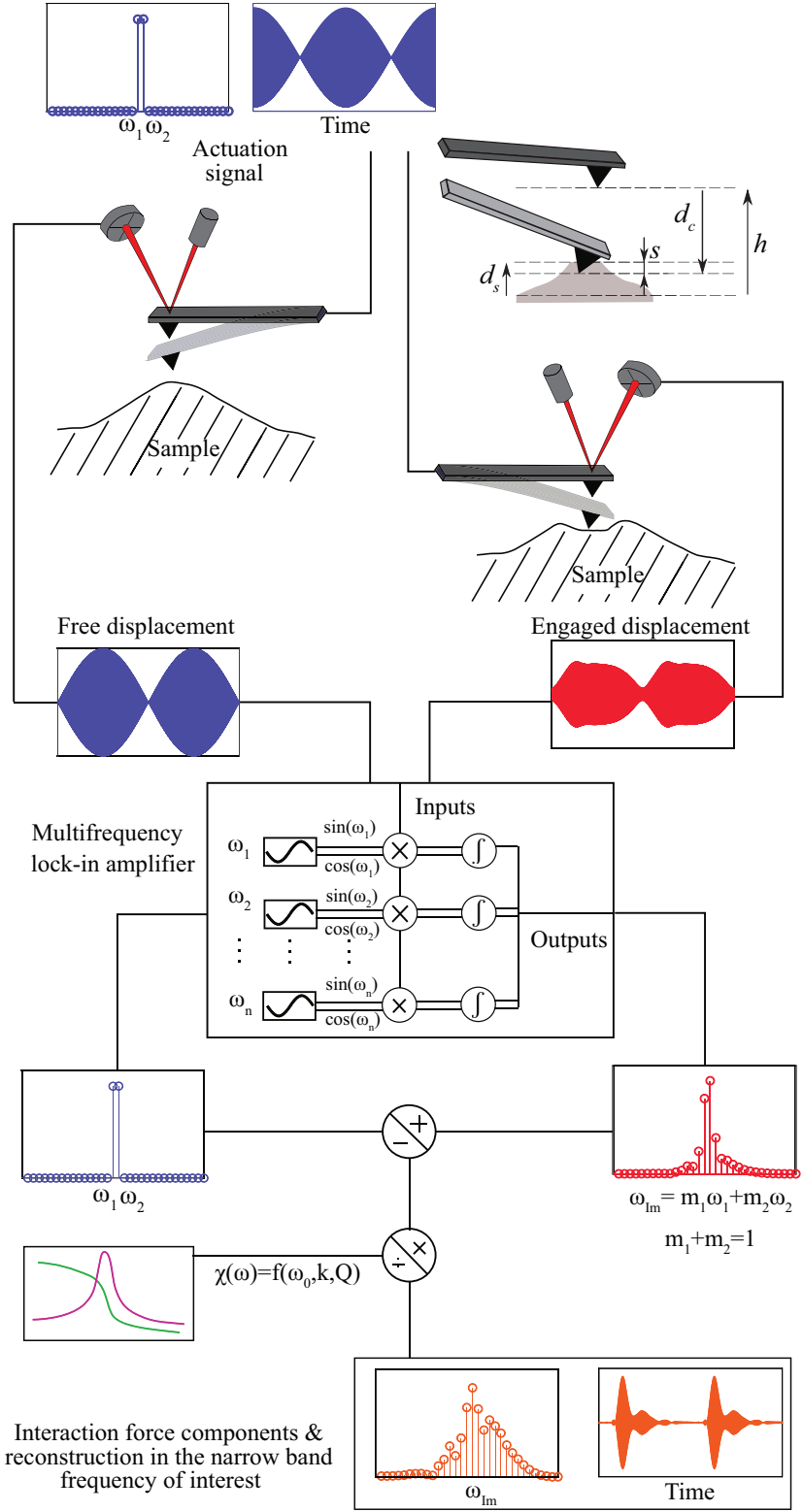}
	\end{minipage}}
	\global\tempheight=\dimexpr \ht0+\dp0+2\intextsep\relax
	\rlap{\usebox0}
\caption{Schematic of the working principle  of the IM-AFM. The cantilever is driven with a signal comprising two close frequencies $\omega_1$ and $\omega_2$, centered around its first resonance frequency. The intermodulation distortion caused by the nonlinear tip-sample interaction creates frequency comb at commensurate frequencies $\omega_{IM} = m_1 \omega_1 + m_2 \omega_2$, with $m_1,m_2 \in \mathrm{Z}$. The linear transfer function of the cantilever $\chi({\omega})$ is measured via thermal calibration, and the amplitudes and phases of these intermodulation products are captured using a multi-lock-in amplifier. Here, $d_c$ and $d_s$ denote the tip cantilever and surface vertical displacements and $h$ corresponds to the unperturbed probe height. Finally, $s=h+d_c-d_s$ represents the tip-sample distance.}
	\label{fig:AFMschematic}
\end{figure}

    We perform our experiments with a commercial AFM (JPK nanowizard 4) and use a multi-lock-in amplifier (Intermodulation products AB) to measure and analyse the frequency components resulting from the tip-sample interaction. A  rectangular
Silicon cantilever (Tap190Al-G, BudgetSensors)   probes the viscoelastic response of a polymer blend made up of PS-LDPE materials. The stiffness of the cantilever ($k$~=~\SI{26.70}{\newton~\per~\meter}), its resonance frequency ($f_0$ = \SI{153.9}{\kilo \hertz }) and the quality factor ($Q$ = 596) are determined using the thermal calibration
method \cite{Higgins2006}. A schematic of the intermodulation AFM setup is shown in Fig.~\ref{fig:AFMschematic}. The cantilever is excited with two frequencies centered around its fundamental mode of vibration. The interaction of the cantilever with the sample, under the influence of nonlinear surface forces, generates frequency combs that are measured using the lock-in amplifier. In particular, the amplitude and phase of the combs are used as experimental inputs for the  viscoelastic identification procedure. Details of IM-AFM operation and processing of the experimental data can be found in \cite{platz2008intermodulation,platz2013interpreting,platz2010phase,haviland2015probing}, we summarize the essential operations in Section S1 of the Supplementary Information (SI). 

\begin{figure}[htbp]
\begin{center}
\includegraphics[width = 12.0cm]{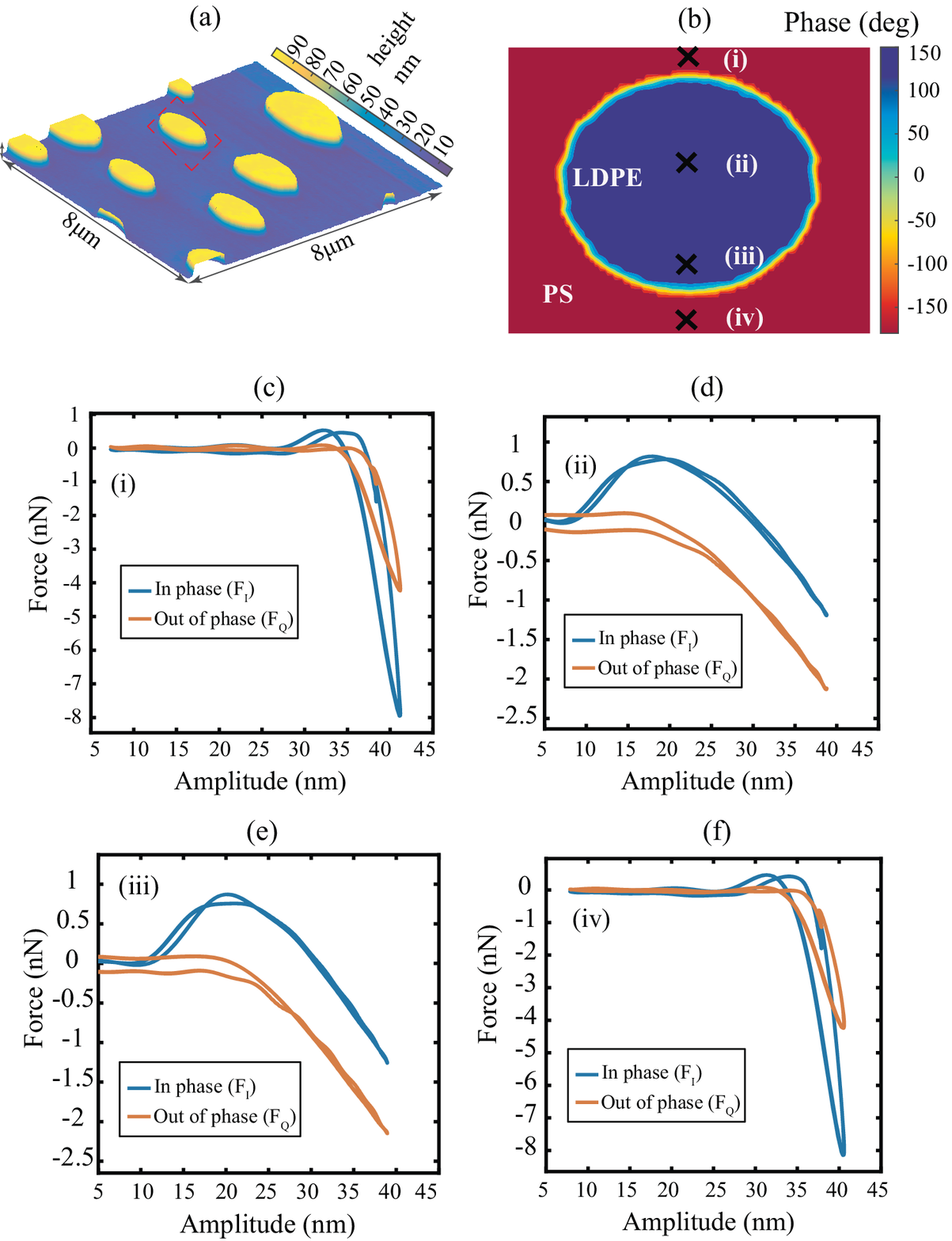}
\end{center}
\caption{Experimental measurements performed on the PS-LDPE polymer blend. (a) Amplitude image at the second drive frequency ($\omega_2$), which is part of the 32 different image pairs captured during the scanning operation. 
(b) Phase image at the second drive frequency. The image shows an island of LDPE within the PS matrix (red dashed box in Fig.~\ref{fig:PSLDPE_measurements_quads}(a)). The points of measurements are indicated with black crosses. (c-f) Experimental force quadratures obtained at the pixels marked by black crosses in the phase image. The quadratures in subfigures (c)-(f) are obtained on PS material, whereas the quadratures in sub figures (d)-(e) are obtained on LDPE material.}
\label{fig:PSLDPE_measurements_quads}
\end{figure}

The experiments performed on the PS-LDPE polymer blend are reported in Fig.~\ref{fig:PSLDPE_measurements_quads}. Figures~\ref{fig:PSLDPE_measurements_quads}(a)-(b) depict the amplitude and phase images at the second drive frequency $\omega_2$. The phase image is presented for one specific LDPE island surrounded by PS matrix. In total 32 amplitude and phase intermodulation components are used to reconstruct the tip-sample interaction in the narrow frequency band around the fundamental resonance. Furthermore, the frequency components are used to calculate the tip-sample force quadratures, which represent the time averaged interaction force that the cantilever experiences in one oscillation cycle (see  Figs.~\ref{fig:PSLDPE_measurements_quads}(c)-(f) for both PS and LDPE). The force quadratures are a local measure of material properties since they are calculated for every pixel of the AFM image; they provide information about the conservative and dissipative contributions of the interaction force between the tip and the sample. For instance, the in-phase quadratures provide  information about the amount of adhesive (positive part) and repulsive (negative part) forces at the measured pixels \cite{haviland2015probing}.


\section{MODELLING TIP-SAMPLE INTERACTION}
 
In order to probe the viscoelastic response of the sample and interpret the in-phase and out-of-phase quadrature information quantitatively, we begin by describing the dynamics of the AFM cantilever using the following simple model \cite{Lee2003,chandrashekar2019robustness}:
\begin{equation}
\frac{1}{\omega_0^2}\ddot{d}_{\text{c}} + \frac{1}{Q \omega_0} \dot{d}_{\text{c}} +d_{\text{c}} = \frac{1}{k} \big(F_{\text{d}}(t) + F_{\text{ts}}(s,\dot{s})\big)\label{eq:CantileverDisplEng},
\end{equation}
where $d_c$ describes the total deflection of the cantilever from its equilibrium,  $\omega_0$=2$\pi f_0$ denotes its resonance frequency, $k$ represents the stiffness of the cantilever, $t$ denotes the time and $F_\text{d}$ is the excitation force. 
 The above equation couples to the sample through the nonlinear tip-surface force 
\begin{equation}
    F_{\text{ts}} (s,\dot{s})=
    \begin{cases}
       -F_{\text{ad}} - k_v s - \eta_v \dot{s}, & \text{if }
s\leq  0, \\
      0 & \text{if } s>0.
 \end{cases}\label{eq:SmapleMot}
  \end{equation} 
 Here, the piecewise linear (PWL) model assumes $F_{\text{ts}}$ to be function of  
  the indentation ($s$) and the rate of indentation ($\dot{s}$).
 In Eq.~\eqref{eq:SmapleMot}, the tip-sample interaction comprises of an adhesion force represented by $F_{\text{ad}}$, a repulsive force due to surface indentation governed by the bulk sample stiffness $k_v$, and finally, a viscous force due to material flow upon indentation governed by the coefficient $\eta_v$.  It must be noted that the PWL model preserves an essential feature of the interaction that is well-known in AFM, which is the presence of large force gradient localized near the point of contact, \textit{i.e} at $s = 0$. 
This rapid change of force is responsible for the jump-to-contact and pull-off hysteresis seen in nearly all quasi-static force curves in AFM. However, in dynamic AFM, the  oscillation amplitude is typically much larger than the range of this localized interaction. Hence, we approximate this region of large interaction gradient as an adhesion force that instantly turns on and off when crossing the point of contact, whose magnitude is counterbalanced by the contribution of the velocity-dependent term $\eta_v \dot{s}$.

We then couple the cantilever dynamics with a moving surface model \cite{haviland2015probing,thoren2018modeling} to account for the motion of the sample interacting with the tip
 
\begin{equation}
\eta_s \dot{d}_s + k_s d_s = -F_{\text{ts}}(s,\dot{s}).
\label{eq:surface_motion}
\end{equation} 
Here, the stiffness and viscosity of the sample  surface are $k_s$ and $\eta_s$, respectively. The instantaneous surface motion is related to the cantilever oscillation through the relation ${s=h+d_{\text{c}}-d_s}$, where $h$ is the unperturbed cantilever height as shown in Fig.~\ref{fig:AFMschematic}.

The tip-sample interaction process as described by Eqs.~\eqref{eq:CantileverDisplEng}-\eqref{eq:surface_motion} introduces a large set of unknown parameters that shall be extracted from the intermodulation components. However, few of them, namely $\omega_0$, $Q$, and $k$ are obtained directly from thermal calibration \cite{sader1999calibration}.  
This reduces the unknown set of parameters that needs to be identified to   ${\bm{P}=\{ F_{\text{ad}}, \enskip k_v, \enskip \eta_v, \enskip k_s, \enskip \eta_s,  \enskip h }\}$. At this stage, the optimization problem is written as:
\begin{equation}
\text{find} \quad 
\text{min}_{\bm{P} \in \mathbb{R}^6} f(\bm{P})\label{eq:DefMinProbl}
\end{equation}with $f(\bm{P})$ the objective function defined as \cite{platz2012role,forchheimer2012model,platz2013interpreting}:
\begin{equation}
f(\bm{P}) = \sqrt{ \sum_{\omega = \omega_{IM}} |\widetilde{F}_{\text{ts,exp}}(\omega) - \widetilde{F}_{\text{ts,sim}}(\omega,\bm{P})|^2 }
\label{eq:fp}
\end{equation}where $\widetilde{F}_{\text{ts,sim}}$ and $\widetilde{F}_{\text{ts,exp}}$ 
denote the complex spectral components of the simulated and experimental interaction force at the intermodulation frequencies $\omega_{IM}$, respectively.

\section{LINKING VISCOELASTICITY TO INTERMODULATION COMPONENTS}\label{sec:Anal_PWLMSM}


We start the identification by analyzing the two pixels denoted by (i) and (iii) in Fig.~\ref{fig:PSLDPE_measurements_quads}(b). These pixels belong to the PS and the LDPE material, respectively. The optimization of the model parameters is carried out using the Levenberg-Marquardt algorithm since it has strong convergence properties and robustness against numerical inconsistencies \cite{ranganathan2004levenberg}. We note that the minima obtained by the optimizer are largely dependent on the initial points (IP) chosen for the unknown parameter set $\bm{P}$. Thus, several initial starting configurations are tested for the identification procedure; these are selected based on values previously reported in the literature \cite{benaglia2019fast,shaik2020nanomechanical,rajabifar2021discrimination,Payam2021} (see Section 3 in SI for additional details).

\begin{table}[htbp]
\centering
  \begin{tabular}{| c | c | c | c | c | c | c |}
  \hline
& \multicolumn{3}{|c|}{Pixel (i) - PS} & \multicolumn{3}{|c|}{Pixel (iii) - LDPE}  \\ \hline 
Initial point & IP 1  & IP 55 & IP 99 & IP 1  &  IP 22 & IP 87 \\ \hline 
$F_{\text{ad}}$ (nN) & 30.5 & 31.6 & 41.6 & 7.08 & 7.12  & 7.13 \\ \hline
 $k_{v}$ (N/m)  & 94.9 & 43.2 &  89.5  &  0.848 & 0.854 &  0.860 \\ \hline 
$\eta_{v}$ (mg/s)  &15.5  & 7.33 &   6.60  &  0.520 & 0.521 & 0.521 \\ \hline 
 $k_{s}$ (N/m)  & 18.8 & 16.8 &  11.8  &  123.8 & 239.3&  28.4\\ \hline
 $\eta_{s}$ (mg/s)  &  0.0552  & 0.00884 & 0.993 &  57.2 & 0.0594 &  62.0 \\ \hline 
$h$ (nm)  &  26.35  & 24.69 &  24.11  &  14.43 & 14.69& 14.67\\ \hline 
 Final $E$ (nN)   &    0.511 & 0.537  & 0.579   & 0.193 & 0.194 & 0.194\\ \hline 
 $R^2$   &  0.961 &  0.957  & 0.950   & 0.979 & 0.979 & 0.979\\  \hline 
  \end{tabular}
    \caption{Extracted results from a large set of local minimization routines using Levenberg-Marquardt algorithm, using the model which includes surface motion and the grid of initial points (IPs) defined in Table S3.5 of SI. The initial points are ranked according to the best results, defined here as the lowest errors / highest $R^2$.}\label{Tab:localopti_results}
\end{table}

\begin{figure}[htbp]
\begin{center}
\includegraphics[width = 10.2cm]{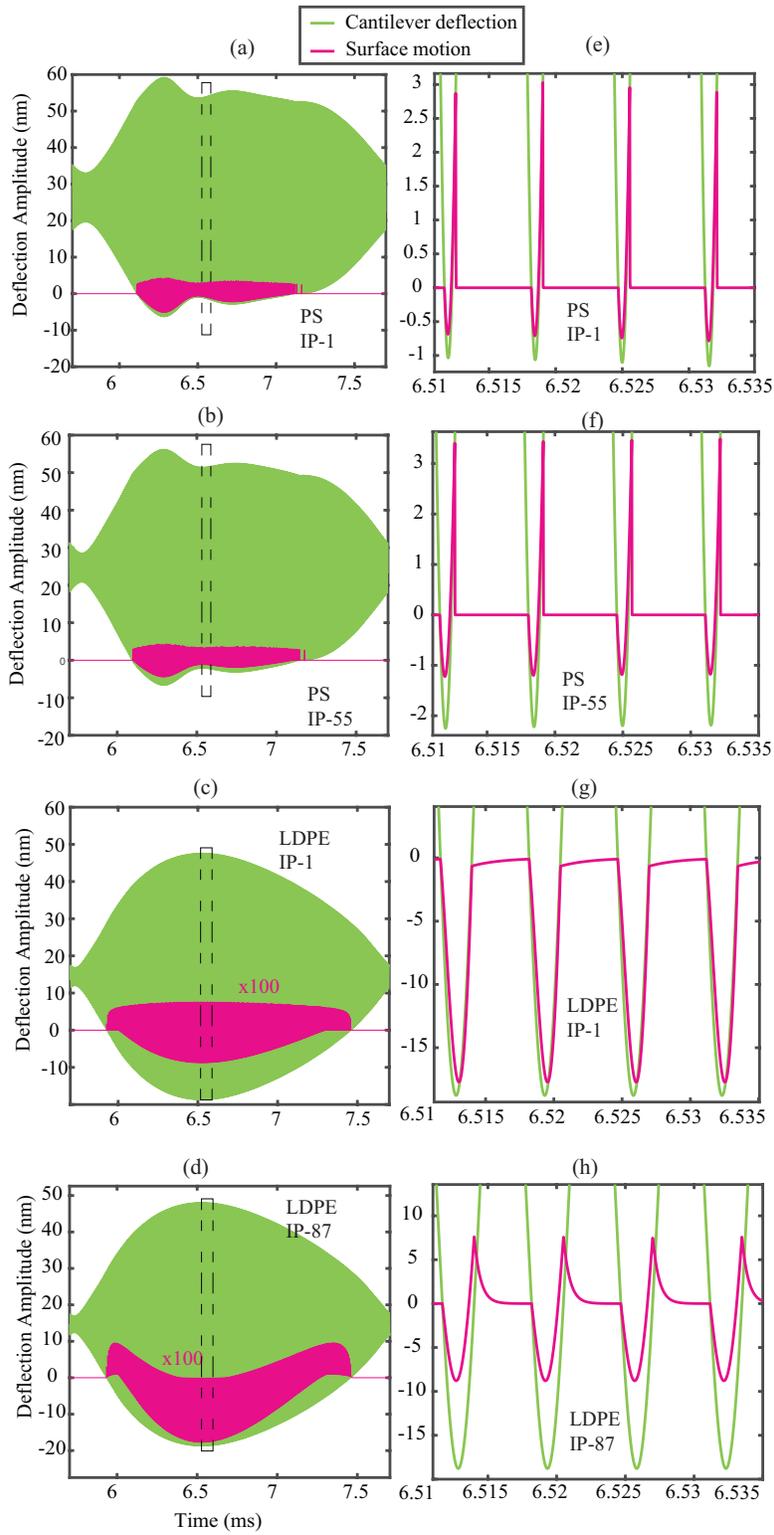}
\end{center}
\caption{Simulations of the cantilever (green) and sample (pink) surface dynamics based on the results provided in table \ref{Tab:localopti_results}. (a)-(b) Simulated results for PS material with parameter values taken from IPs 1 and 55, respectively. (c)-(d) Simulated results for LDPE material with parameter values taken from IPs 1 and 87. (e)-(f)-(g)-(h) A close up visualization of the surface dynamics is reported in (a)-(b)-(c)-(d).}
\label{fig:local_min_surf_motion}
\end{figure}

Table~\ref{Tab:localopti_results}  summarises the identified model parameters and the corresponding errors between the simulation and the experimental counterparts for several different IPs on pixels (i) and (iii). Here, we note that the surface stiffness ($k_s$) and damping ($\eta_s$) of LDPE is much higher than PS matrix which is intuitively wrong since PS is the stiffer material. In addition to this, we observe from Fig.~\ref{fig:local_min_surf_motion}(a)-(b) that the reconstructed cantilever motion (green) and the surface motion (pink) look identical, even though they represent different set of identified values (See Table~\ref{Tab:localopti_results}). Moreover, in Fig.~\ref{fig:local_min_surf_motion}(c)-(d) the surface motion in case of LDPE is strongly dependent on the choice of IPs and consequently leads to different parameter value estimations. Contrary to the popular notion, the amplitude of surface motion in case of soft LDPE is also much smaller when compared to the stiff PS material.

\begin{figure*}[htbp]
	\sbox0{
		\begin{minipage}{\columnwidth}
			\centering
			\includegraphics[width=0.9\columnwidth]{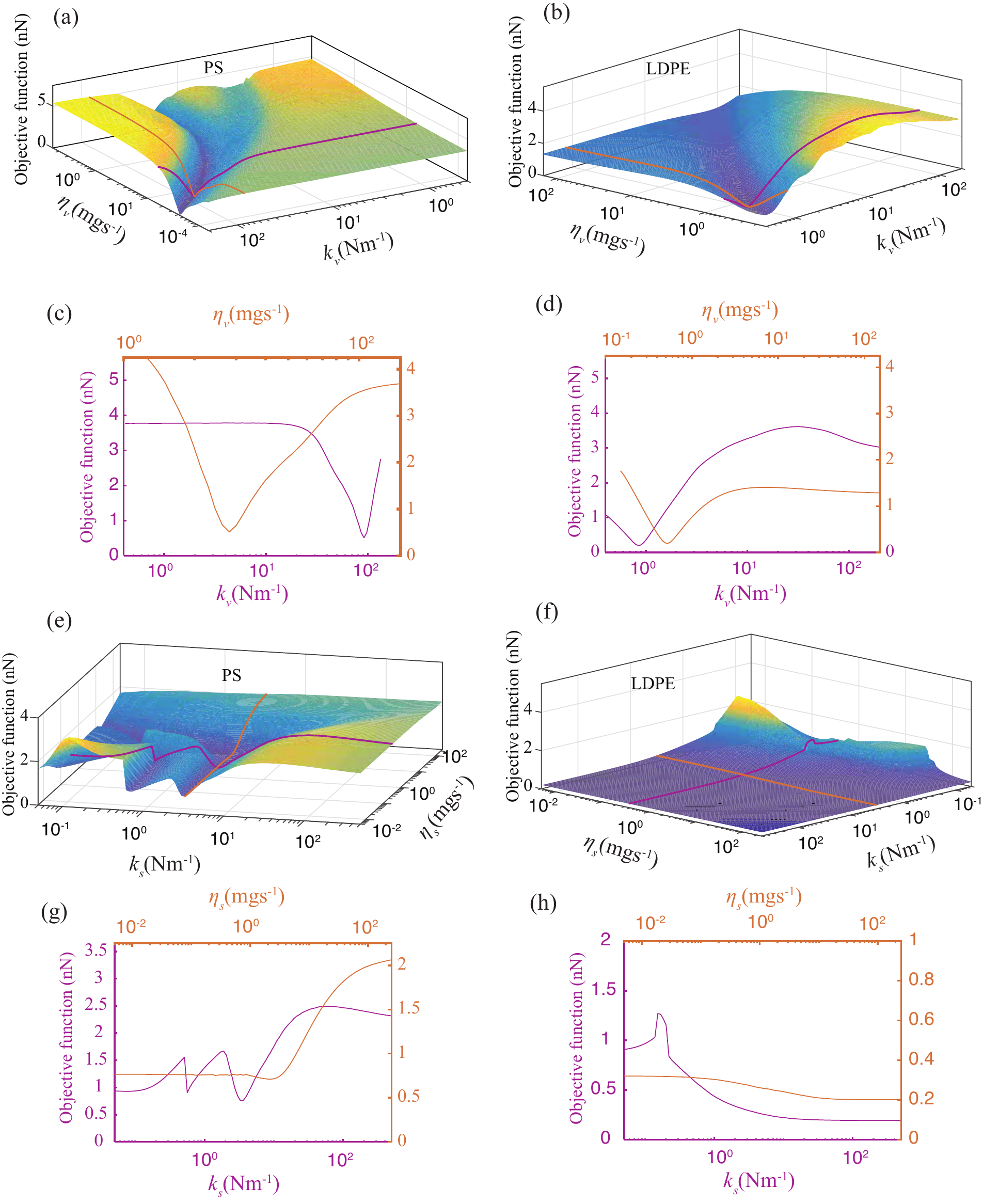}
	\end{minipage}}
	\global\tempheight=\dimexpr \ht0+\dp0+2\intextsep\relax
	\rlap{\usebox0}
	\caption{Variation of the objective function in a 2-dimensional parameter space comprising ($(k_{s},\eta_{s})$ or $(k_{v},\eta_{v})$), with the other parameters fixed in accordance with the best results found from the local minimization routine. (a)-(d) Visualizing the landscape  of the minimization objective as a function of $k_v$ and $\eta_v$ for PS and LDPE material obtained at pixel (i) and (iii) of Fig.~\ref{fig:PSLDPE_measurements_quads}(b). The Pink and orange lines indicate a 2D cross-sectional view of the objective function. (a)-(d) Visualizing the landscape of the minimization objective as a function of $k_s$ and $\eta_s$ for PS and LDPE material obtained at pixel (i) and (iii) of Fig.~\ref{fig:PSLDPE_measurements_quads}(b). Pink and orange lines indicate  2D cross-sectional views of the objective function. }\label{fig:obj_func}
\end{figure*}

We relate the above discrepancies to possible  insensitivity of the objective function towards certain model parameters and the presence of multiple local minima, which indicates that the objective function is non-convex. To elaborate on these issues, we analyze the topological landscape of the objective function on a larger parameter range. We note that the objective function includes 6 parameters, out of which $F_{\text{ad}}$ and $h$ show consistent convergence. Hence, we limit our analysis to the bulk and surface viscoelastic parameters governed by $k_v$, $\eta_v$, $k_s$, and $\eta_s$. This is showcased in Fig.~\ref{fig:obj_func}, where  topological landscapes of the objective function are obtained by sweeping across the viscoelastic parameters for both PS and LDPE material at pixels (i) and  (iii), respectively. In each sub-figure, the four non-varied parameters are chosen as those of IP 1 in Table~\ref{Tab:localopti_results}. Interestingly, we note that Figs.~\ref{fig:obj_func}(a)-(b) exhibit a valley in which a single optimum solution is found. This is further highlighted in the 2D cross sections shown as Figs.~\ref{fig:obj_func}(c)-(d) , confirming the strong dependency of parameters $k_v$ and $\eta_v$ on the experimental observables. Contrary to this, the objective landscape of Figs.~\ref{fig:obj_func}(e)-(f) highlight multiple local minima (in the case of pixel (i) in Fig.~\ref{fig:obj_func}(e)) or a flat insensitive gradient (for pixel (iii), in Fig.~\ref{fig:obj_func}(f)). A flat landscape of the objective function in case of softer LDPE is counter-intuitive since one would expect a softer material to show pronounced surface dynamics compared to PS. This behaviour is also reflected in the large spread of values reported in Table~\ref{Tab:localopti_results}.

In order to verify that the  discrepancy does not stem from the optimizer used, we also employ a heuristic global optimization technique in pursuit of a global solution in the parameter space.  We create synthetic data sets with known optima to analyse how the global optimizer performs (for details see Section 2.2 in SI). Once again the optimizer fails to overcome the aforementioned discrepancies. Since a wide range of non-physical parameter values reconstructs the cantilever motion while  surface viscoelastic parameters do not affect  the objective function. Upon closer inspection of results, we noticed a trend for synthetic data sets with good solution convergence, where the bulk parameters of the model, namely $k_v$, $\eta_v$, tends to the original optimum (for details see table S2.4 in Section 2.2 of SI). This is in  accordance with our hypothesis regarding the insensitivity of surface viscoelastic parameters on the experimental observables. Therefore, fine-tuning of the global optimization parameter space is effective in determining bulk viscoelastic parameters. Nevertheless, isolation of non-physical solutions as outliers is computationally expensive when aiming for fast parameter estimation.
For this reason we explore an alternative local optimization route paired with an initial point selection procedure in the following section.

\subsection*{Estimating bulk viscoelasticity in the absence of surface motion}
In order to overcome the aforementioned limitations as well as to improve the computational efficiency for the parameter estimation procedure, we neglect the surface dynamics of the sample and reduce the unknown parameter set to   $\bm{\bar{P}}=\{ F_{\text{ad}}, \enskip k_v, \enskip \eta_v, \enskip h\}$. It must be noted that this reduced set is still descriptive of the nanomechanical mapping of polymer blends and coherent with several  well-established formulations, \textit{e.g.}, Derjaguin-Muller-Toporov (DMT)-Kelvin-Voigt \cite{benaglia2019fast}, 3D Kelvin-Voigt \cite{Garcia2018},  and DMT-Garcia \cite{garcia2006identification}.

\begin{figure}[htbp]
	\sbox0{
		\begin{minipage}{\columnwidth}
			\centering
			\includegraphics[width=0.65\columnwidth]{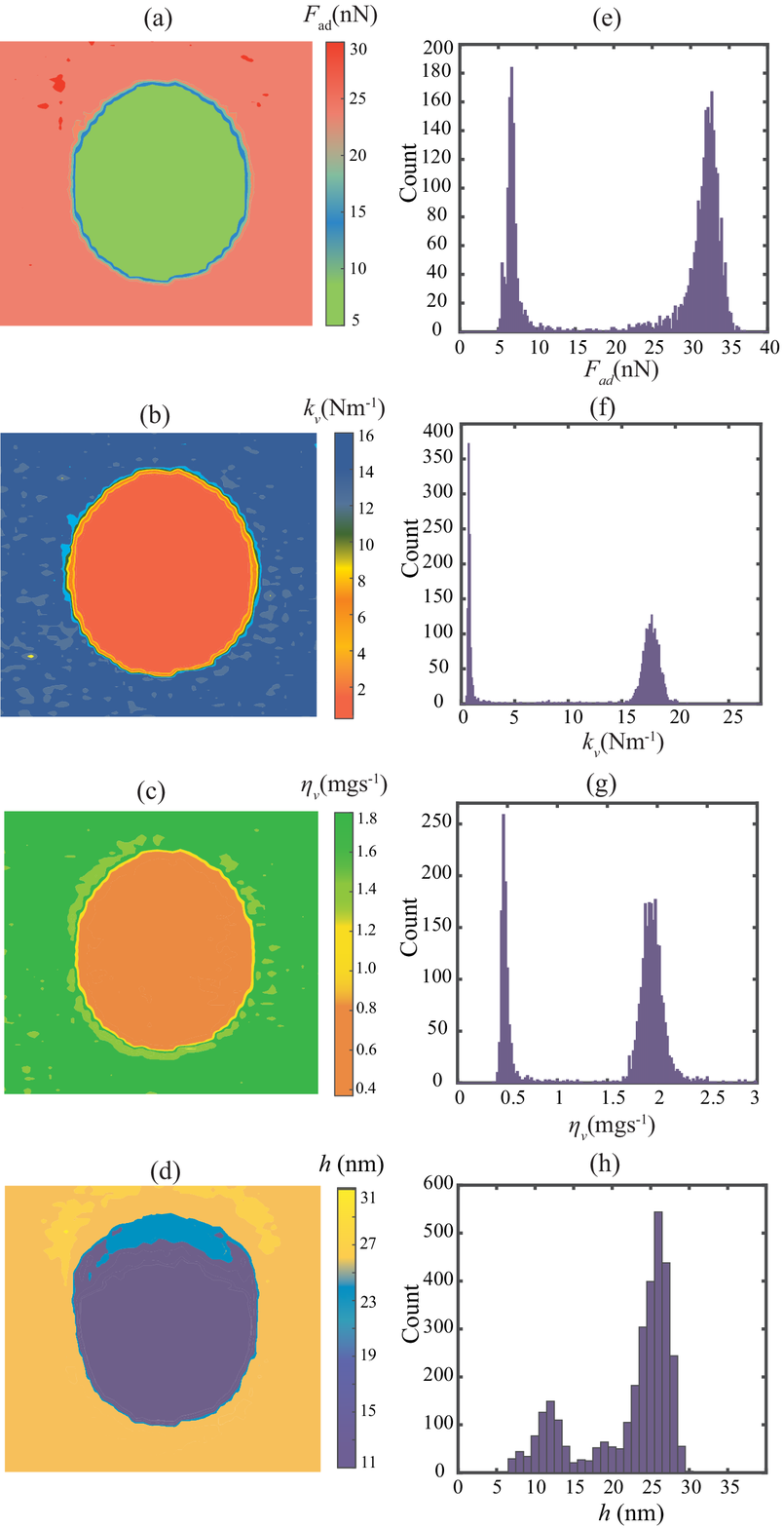}
	\end{minipage}}
	\global\tempheight=\dimexpr \ht0+\dp0+2\intextsep\relax
	\rlap{\usebox0}
	\caption{Estimated properties of the PS-LDPE sample obtained using the model without surface motion and the initial points selection procedure described in this work. The maps are of dimensions \SI{2.5}{\micro \meter} x \SI{2.5}{\micro \meter}. Left: parameter maps of (a) adhesion force, (b) contact force stiffness, (c) contact force viscosity, (d) Probe height. Right: histogram distribution of the respective parameters (e-f-g-h).}\label{fig:PSldpe_est_prop}
\end{figure}

We begin by repeating the quantitative analysis at pixels (i) and (iii) of Fig.~\ref{fig:PSLDPE_measurements_quads}, once again applying the Levenberg-Marquardt algorithm. In this procedure we use a grid of $3^4$ IPs, by defining three values for the four free parameters of the model. This choice of three values is motivated by a compromise between a wide range of parameter exploration and a reasonable simulation duration. These parameter values include in particular at least one order of magnitude for the viscoelastic properties (for details see Section S3.2 in SI). Furthermore, the three values of the probe height $h$ can be framed from the force quadrature profiles and from onsets of repulsive forces (for details see Section S2.3 in SI).
We then perform a gradient-based optimization for each combination of parameters in the parameter space and conduct statistical analysis by obtaining the Gaussian distribution profiles of the identified parameters (for more details see Section S3.2 in SI). Interestingly, for most of the IPs  the  optimizer converges towards an admissible physical solution.

Based on this statistical analysis we extract a set of three initial points for performing the parameter identification at all pixels of the entire AFM scan. The first two sets of IPs are derived from the mean values of the Gaussian distribution for both the PS and LDPE material. Indeed, these mean values lead to the lowest errors at pixels (i) and (iii). As for the third set, an IP is chosen which can lead to a set of identified parameter within a specific confidence interval for both the PS and LDPE material. The reasoning for choosing such an IP is rooted in our optimization procedure where, we assume that pixels belonging to the same material have similar objective function topology. This assumption may not hold true at the junctions where the two materials blend. Hence, having a third IP that could identify the parameters of both PS and LDPE material within a certain confidence interval is crucial to avoid non-physical parameter estimation (for details see section 3.2 of SI). Finally, among the three optimization run at each pixel, we retain the parameters of the best fit (\textit{i.e.} the lowest error) as the identified model parameters.

Figure~\ref{fig:PSldpe_est_prop} shows the identified parameter values for the PS-LDPE polymer blend. It highlights a clear distinction between the identified bulk parameters $F_{\text{ad}}$, $k_v$, and $\eta_v$ between the island of LDPE and the surrounding PS matrix. This can be seen in the observed compositional contrast in the colored figures. Additionally, the histogram displayed on the right side of the figure highlights clear separated Gaussian profiles for each of the parameters. The estimated values lie within a 95\% confidence interval for the entire image, as Table.~\ref{Tab:without_surf_results} shows. Moreover, we remark that our identified values are in line with those previously reported in the literature \cite{benaglia2019fast,shaik2020nanomechanical,rajabifar2021discrimination,Payam2021} and  align  with the expected physical behaviour of the two polymers, i.e ($F_{\text{a,PS}} > F_{\text{a,LDPE}}$, $k_{\text{v,PS}} > k_{\text{v,LDPE}}$ and $\eta_{\text{v,PS}} > \eta_{\text{v,LDPE}}$). Our analysis suggests that intermodulation frequency components have a direct correlation with the bulk properties of the sample and the interaction force function can be robustly characterized.
\begin{table}[ht!]
\centering
  \begin{tabular}{| c | c | c |}
  \hline
 & PS  & LDPE \\ \hline 
$F_{\text{ad}}$ (nN) & $31.49 \pm  0.12 $ & $6.960\pm  0.076$ \\ \hline
 $k_{v}$ (N/m)  & $17.31 \pm  0.09$ & $0.819\pm  0.020$ \\ \hline
$\eta_{v}$ (mg/s)  & $1.951\pm  0.007$ & $0.492 \pm  0.005$ \\ \hline
$h$ (nm)  & $26.71 \pm  0.02 $  &  $12.86 \pm 0.021 $     \\ \hline
  \end{tabular}
    \caption{Identified parameters resulting from the Gaussian fits, made from the material properties estimated at all pixels plotted in Fig.~\ref{fig:PSldpe_est_prop}. The uncertainties are estimated with a 95\% confidence interval.
    }\label{Tab:without_surf_results}
\end{table}
 
\section{Conclusions}

In summary, we studied the dependency of viscoelastic response of polymeric samples to multi-frequency IM-AFM. We discussed the sensitivity issues that can be faced when minimizing the error between IM-AFM spectral components and a tip-sample force model with surface dynamics, and confirmed that insensitivity of surface viscoelasticity to experimental observables could lead to non-physical parameter estimations. We attribute this finding to the non-convexity and flat topological landscape of the objective function with respect to the sample's surface viscoelastic parameters. This was further reinforced with numerical simulations that used both gradient-based and heuristic global optimization techniques. We remedy this issue with a simplified model that only accounts for the bulk viscoelastic parameters and by implementing an initial point selection procedure that searches a large parameter space to estimate model unknowns with ease. This new framework results in consistent identification of viscoelastic parameters that are in good agreement with  previously reported values. However, in order to take full advantage of the vast amount of multi-frequency observables, a more accurate and sensitive viscoelastic tip-surface model is needed \cite{Rajabifar_2021,rajabifar2021discrimination,Ting1966}, and computational developments to speed up the optimization process are required.
 Finally, given the growing interest in developing multi-parametric  techniques in multi-frequency AFM, we believe that the techniques showcased in this work can be useful in providing guidance to future investigations that are aimed at studying soft, adhesive and viscoelastic surfaces of samples. 
 

\noindent
\section*{Author Contributions}
A.C, C.P, and F.A conceived the experiments. A.C and A.G prepared the samples and 
conducted the experiments. A.C, A.G, P.B, and FA conceived the simulations. A.G and C.P conducted the simulations. A.C, A.G, P.B, C.P, A.A and F.A did data analysis and interpretation. U.S and F.A supervised the project. All the authors jointly wrote the article with main contribution from A.C and A.G. All authors discussed the results and commented on the article.\\

\section*{Conflicts of interest}
 The authors declare no competing interests.

\section*{Acknowledgements} 
This work is part of the research programme
’A NICE TIP TAP’ with grant number 15450 which is financed by the
Netherlands Organisation for Scientific Research (NWO). FA also acknowledges financial support from  European Union’s Horizon 2020 research and innovation program under Grant Agreement 802093 (ERC starting grant ENIGMA).\\

 \section*{Data availability}
The authors declare that all the data in this manuscript are available upon request.\\

\bibliographystyle{unsrtPB}%
\bibliography{biblio}

\begin{thebibliography}{10}

\bibitem{raman2011mapping}
A.~Raman, S.~Trigueros, A.~Cartagena, A.~Stevenson, M.~Susilo, E.~Nauman, and
  S.~A. Contera.
\newblock Mapping nanomechanical properties of live cells using multi-harmonic
  atomic force microscopy.
\newblock {\em Nature nanotechnology}, 6(12):809--814, 2011.

\bibitem{Uhlig2017}
M.~R. Uhlig and R.~Magerle.
\newblock Unraveling capillary interaction and viscoelastic response in atomic
  force microscopy of hydrated collagen fibrils.
\newblock {\em Nanoscale}, 9(3):1244--1256, 2017.

\bibitem{efremov2017measuring}
Y.~M. Efremov, W.-H. Wang, S.~D. Hardy, R.~L. Geahlen, and A.~Raman.
\newblock Measuring nanoscale viscoelastic parameters of cells directly from
  afm force-displacement curves.
\newblock {\em Scientific reports}, 7(1):1--14, 2017.

\bibitem{efremov2020measuring}
Y.~Efremov, T.~Okajima, and A.~Raman.
\newblock Measuring viscoelasticity of soft biological samples using atomic
  force microscopy.
\newblock {\em Soft matter}, 16(1):64--81, 2020.

\bibitem{Chiefari1998}
J.~Chiefari, Y.~K.~B. Chong, F.~Ercole, J.~Krstina, J.~Jeffery, T.~P.~T. Le,
  R.~T.~A. Mayadunne, G.~F. Meijs, C.~L. Moad, G.~Moad, E.~Rizzardo, and S.~H.
  Thang.
\newblock Living free-radical polymerization by reversible
  addition-fragmentation chain transfer: The raft process.
\newblock {\em Macromolecules}, 31(16):5559--5562, Aug 1998.

\bibitem{Chyasnavichyus2014}
M.~Chyasnavichyus, S.~L. Young, and V.~V. Tsukruk.
\newblock Probing of polymer surfaces in the viscoelastic regime.
\newblock {\em Langmuir}, 30(35):10566--10582, 2014.

\bibitem{proksch2016practical}
R.~Proksch, M.~Kocun, D.~Hurley, M.~Viani, A.~Labuda, W.~Meinhold, and
  J.~Bemis.
\newblock Practical loss tangent imaging with amplitude-modulated atomic force
  microscopy.
\newblock {\em Journal of Applied Physics}, 119(13):134901, 2016.

\bibitem{Garcia2014}
R.~Garcia, A.~W. Knoll, and E.~Riedo.
\newblock Advanced scanning probe lithography.
\newblock {\em Nature Nanotechnology}, 9(8):577--587, 2014.

\bibitem{Wang2018}
D.~Wang and T.~P. Russell.
\newblock Advances in atomic force microscopy for probing polymer structure and
  properties.
\newblock {\em Macromolecules}, 51(1):3--24, 2018.

\bibitem{garcia2020nanomechanical}
R.~Garcia.
\newblock Nanomechanical mapping of soft materials with the atomic force
  microscope: methods, theory and applications.
\newblock {\em Chemical Society Reviews}, 49(16):5850--5884, 2020.

\bibitem{collinson2021best}
D.~W. Collinson, R.~J. Sheridan, M.~J. Palmeri, and L.~C. Brinson.
\newblock Best practices and recommendations for accurate nanomechanical
  characterization of heterogeneous polymer systems with atomic force
  microscopy.
\newblock {\em Progress in Polymer Science}, 119:101420, 2021.

\bibitem{platz2008intermodulation}
D.~Platz, E.~Thol{\'e}n, D.~Pesen, and D.~Haviland.
\newblock Intermodulation atomic force microscopy.
\newblock {\em Applied Physics Letters}, 92(15):153106, 2008.

\bibitem{platz2013interpreting}
D.~Platz, D.~Forchheimer, E.~A. Thol{\'e}n, and D.~B. Haviland.
\newblock Interpreting motion and force for narrow-band intermodulation atomic
  force microscopy.
\newblock {\em Beilstein journal of nanotechnology}, 4(1):45--56, 2013.

\bibitem{platz2013interaction}
D.~Platz, D.~Forchheimer, E.~A. Thol{\'e}n, and D.~B. Haviland.
\newblock Interaction imaging with amplitude-dependence force spectroscopy.
\newblock {\em Nature communications}, 4(1):1--9, 2013.

\bibitem{Ting1966}
T.~C.~T. Ting.
\newblock {The Contact Stresses Between a Rigid Indenter and a Viscoelastic
  Half-Space}.
\newblock {\em Journal of Applied Mechanics}, 33(4):845--854, 12 1966.

\bibitem{attard2001interaction}
P.~Attard.
\newblock Interaction and deformation of viscoelastic particles. 2. adhesive
  particles.
\newblock {\em Langmuir}, 17(14):4322--4328, 2001.

\bibitem{rajabifar2018dynamic}
B.~Rajabifar, J.~M. Jadhav, D.~Kiracofe, G.~F. Meyers, and A.~Raman.
\newblock Dynamic afm on viscoelastic polymer samples with surface forces.
\newblock {\em Macromolecules}, 51(23):9649--9661, 2018.

\bibitem{solares2014probing}
S.~D. Solares.
\newblock Probing viscoelastic surfaces with bimodal tapping-mode atomic force
  microscopy: Underlying physics and observables for a standard linear solid
  model.
\newblock {\em Beilstein journal of nanotechnology}, 5(1):1649--1663, 2014.

\bibitem{thoren2018modeling}
P.-A. Thor{\'e}n, R.~Borgani, D.~Forchheimer, I.~Dobryden, P.~Claesson,
  H.~Kassa, P.~Lecl{\`e}re, Y.~Wang, H.~Jaeger, and D.~Haviland.
\newblock Modeling and measuring viscoelasticity with dynamic atomic force
  microscopy.
\newblock {\em Physical Review Applied}, 10(2):024017, 2018.

\bibitem{Rajabifar_2021}
B.~Rajabifar, R.~Wagner, and A.~Raman.
\newblock A fast first-principles approach to model atomic force microscopy on
  soft, adhesive, and viscoelastic surfaces.
\newblock {\em Materials Research Express}, 8(9):095304, sep 2021.

\bibitem{haviland2015probing}
D.~B. Haviland, C.~A. van Eysden, D.~Forchheimer, D.~Platz, H.~G. Kassa, and
  P.~Lecl{\`e}re.
\newblock Probing viscoelastic response of soft material surfaces at the
  nanoscale.
\newblock {\em Soft Matter}, 12(2):619--624, 2015.

\bibitem{Higgins2006}
M.~J. Higgins, R.~Proksch, J.~E. Sader, M.~Polcik, S.~Mc~Endoo, J.~P.
  Cleveland, and S.~P. Jarvis.
\newblock Noninvasive determination of optical lever sensitivity in atomic
  force microscopy.
\newblock {\em Review of Scientific Instruments}, 77(1):013701, 2006.

\bibitem{platz2010phase}
D.~Platz, E.~A. Thol{\'e}n, C.~Hutter, A.~C. von Bieren, and D.~B. Haviland.
\newblock Phase imaging with intermodulation atomic force microscopy.
\newblock {\em Ultramicroscopy}, 110(6):573--577, 2010.

\bibitem{Lee2003}
S.~I. Lee, S.~W. Howell, A.~Raman, and R.~Reifenberger.
\newblock Nonlinear dynamic perspectives on dynamic force microscopy.
\newblock {\em Ultramicroscopy}, 97(1-4):185--198, 2003.

\bibitem{chandrashekar2019robustness}
A.~Chandrashekar, P.~Belardinelli, U.~Staufer, and F.~Alijani.
\newblock Robustness of attractors in tapping mode atomic force microscopy.
\newblock {\em Nonlinear Dynamics}, 97(2):1137--1158, 2019.

\bibitem{sader1999calibration}
J.~E. Sader, J.~W. Chon, and P.~Mulvaney.
\newblock Calibration of rectangular atomic force microscope cantilevers.
\newblock {\em Review of scientific instruments}, 70(10):3967--3969, 1999.

\bibitem{platz2012role}
D.~Platz, D.~Forchheimer, E.~A. Thol{\'e}n, and D.~B. Haviland.
\newblock The role of nonlinear dynamics in quantitative atomic force
  microscopy.
\newblock {\em Nanotechnology}, 23(26):265705, 2012.

\bibitem{forchheimer2012model}
D.~Forchheimer, D.~Platz, E.~A. Thol{\'e}n, and D.~B. Haviland.
\newblock Model-based extraction of material properties in multifrequency
  atomic force microscopy.
\newblock {\em Physical Review B}, 85(19):195449, 2012.

\bibitem{ranganathan2004levenberg}
A.~Ranganathan.
\newblock The levenberg-marquardt algorithm.
\newblock {\em Tutoral on LM algorithm}, 11(1):101--110, 2004.

\bibitem{benaglia2019fast}
S.~Benaglia, C.~A. Amo, and R.~Garcia.
\newblock Fast, quantitative and high resolution mapping of viscoelastic
  properties with bimodal afm.
\newblock {\em Nanoscale}, 11(32):15289--15297, 2019.

\bibitem{shaik2020nanomechanical}
N.~H. Shaik, R.~G. Reifenberger, and A.~Raman.
\newblock Nanomechanical mapping in air or vacuum using multi-harmonic signals
  in tapping mode atomic force microscopy.
\newblock {\em Nanotechnology}, 31(45):455502, 2020.

\bibitem{rajabifar2021discrimination}
B.~Rajabifar, A.~K. Bajaj, R.~G. Reifenberger, R.~Proksch, and A.~Raman.
\newblock Discrimination of adhesion and viscoelasticity from nanoscale maps of
  polymer surfaces using bimodal atomic force microscopy.
\newblock {\em Nanoscale}, 2021.

\bibitem{Payam2021}
A.~F. Payam, A.~Morelli, and P.~Lemoine.
\newblock Multiparametric analytical quantification of materials at nanoscale
  in tapping force microscopy.
\newblock {\em Applied Surface Science}, 536:147698, 2021.

\bibitem{Garcia2018}
P.~D. Garcia and R.~Garcia.
\newblock Determination of the viscoelastic properties of a single cell
  cultured on a rigid support by force microscopy.
\newblock {\em Nanoscale}, 10:19799--19809, 2018.

\bibitem{garcia2006identification}
R.~Garcia, C.~Gomez, N.~Martinez, S.~Patil, C.~Dietz, and R.~Magerle.
\newblock Identification of nanoscale dissipation processes by dynamic atomic
  force microscopy.
\newblock {\em Physical review letters}, 97(1):016103, 2006.

\end{thebibliography}


\begin{thebibliography}{10}
\bibitem[1]{borgani2017backgroundSup} R.~Borgani, P.-A.~Thorén, D.~Forchheimer, I.~Dobryden, S.~M.~Sah, P.~M.~Claesson and D.~B.~Haviland. Background-force compensation in dynamic atomic force microscopy. \textit{Physical Review Applied}, 7(6):064018, 2017.
\bibitem[2]{hutter2010reconstructingSup} C.~Hutter, D.~Platz, E.~A.~Tholén, T.~Hansson and D.~B.~Haviland. Reconstructing nonlinearities with intermodulation spectroscopy, \textit{Physical Review Letters} 104(5):050801, 2010.
\bibitem[3]{platz2013reconstructingSup} D.~Platz. \textit{Reconstructing force from harmonic motion.} Ph.D. Thesis, KTH Royal Institute of Technology, 2013.
\bibitem[4]{platz2013interpretingSup} D.~Platz, D.~Forchheimer, E.~A.~Tholén and D.~B.~Haviland. Interpreting motion and force for narrow-band intermodulation atomic force microscopy. \textit{Beilstein journal of nanotechnology}, 4(1):45–56, 2013.
\bibitem[5]{tholen2011noteSup} E.~A.~Tholén, D.~Platz, D.~Forchheimer, V.~Schuler, M.~O.~Tholén, C.~Hutter and D.~B.~Haviland. Note: The intermodulation lockin analyzer. \textit{Review of Scientific Instruments} 82(2):026109, 2011.
\bibitem[6]{forchheimer2012modelSup} D.~Forchheimer, D.~Platz, E.~A.~Tholén and D.~B.~Haviland. Model-based extraction
of material properties in multifrequency atomic force microscopy. \textit{Physical Review B},
85(19):195449, 2012.
\bibitem[7]{levenberg1944methodSup} K.~Levenberg. A method for the solution of certain non-linear problems in least squares, \textit{Quarterly of applied mathematics}, 2(2):164--168, 1944.
\bibitem[8]{Thoren2018Sup} P.-A.~Thorén, R.~Borgani, D.~Forchheimer, I.~Dobryden, P.~Claesson, H.~Kassa, P.~Leclère, Y.~Wang, H.~Jaeger and D.~Haviland. Modeling and measuring viscoelasticity with dynamic atomic force microscopy. \textit{Physical Review Applied}, 10(2):024017,
2018.
\bibitem[9]{haviland2015probingSup} D.~B.~Haviland, C.~A.~van Eysden, D.~Forchheimer, D.~Platz, H.~G.~Kassa and
P.~Leclère. Probing viscoelastic response of soft material surfaces at the nanoscale.
\textit{Soft Matter}, 12(2):619–624, 2015.
\bibitem[10]{Penning2020Sup} C.~L.~Penning, Modelling of viscoelasticity using multifrequency {AFM}, 2020.
\bibitem[11]{attard2001interactionSup} P.~Attard. Interaction and deformation of viscoelastic particles. 2. Adhesive particles, \textit{Langmuir} 17(14):4322, 2001.
\bibitem[12]{attard2007measurementSup} P.~Attard. Measurement and interpretation of elastic and viscoelastic properties with
the atomic force microscope. \textit{Journal of Physics: Condensed Matter}, 19(47):473201,
2007.
\bibitem[13]{rajabifar2018dynamicSup} B.~Rajabifar, J.~M.~Jadhav, D.~Kiracofe, G.~F.~Meyers and A.~Raman. Dynamic {AFM}
on viscoelastic polymer samples with surface forces. \textit{Macromolecules}, 51(23):9649–9661, 2018.
\bibitem[14]{rajabifar2021discriminationSup} B.~Rajabifar, A.~K.~Bajaj, R.~G.~Reifenberger, R.~Proksch and A.~Raman. Discrim-
ination of adhesion and viscoelasticity from nanoscale maps of polymer surfaces using
bimodal atomic force microscopy. \textit{Nanoscale} 13(41):17428--17441, 2021.
\bibitem[15]{rajabifar2021fastSup} B.~Rajabifar, R.~Wagner and A.~Raman. A fast first-principles approach to model atomic force microscopy on soft, adhesive, and viscoelastic surfaces. \textit{Materials Research Express}, 8(9):095304, 2021.
\end{thebibliography}

\newpage

\title{: Supporting Information}

\begin{frontmatter}

\end{frontmatter}

\renewcommand{\thesection}{S\arabic{section}}  
\renewcommand{\thetable}{S\arabic{table}}  
\renewcommand{\thefigure}{S\arabic{figure}} 
\setcounter{section}{0}
\section{Experimental data processing}

\begin{figure}[htbp]
\begin{small}
\begin{center}\def\svgwidth{16.8cm}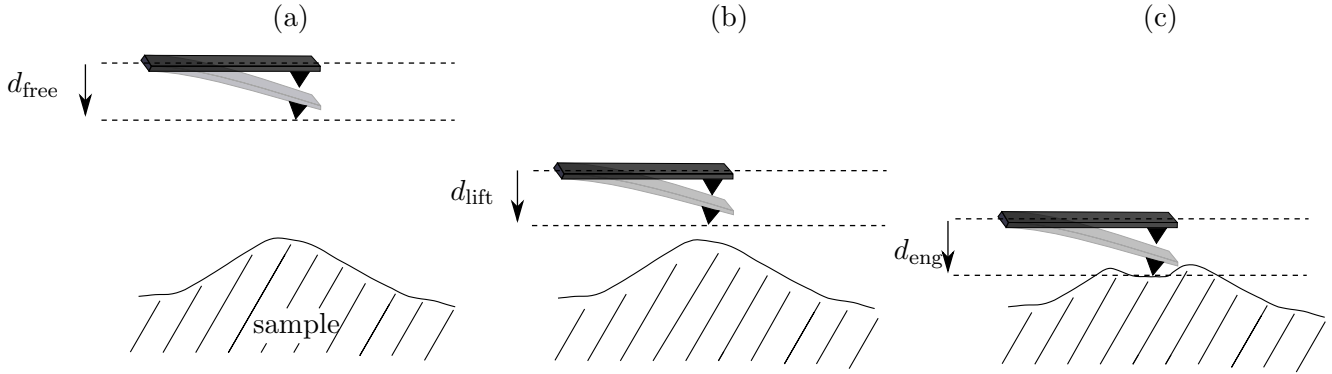
\end{center}\caption{Schematic of the different working positions ((a) free - (b) lift - (c) engaged) for the measurements of the cantilever displacement.}
\label{fig:ScheLift}
\end{small}
\end{figure}

We measure the spectral components of the cantilever motions $d_c$ in free, lift and engaged 
motions $(\widetilde{d}_{\text{free}},\widetilde{d}_{\text{eng}}, \widetilde{d}_{\text{lift}})$, which correspond to tip motions measured at decreasing distances from the sample as described in Fig.~\ref{fig:ScheLift}.\\

 The lift motion denotes the motion of the cantilever at a position close to the surface. It provides a measure to compensate the contribution of long-range linear forces due  to squeeze-film damping or electrostatic interactions. These effects are embedded in the linear transfer function of the so-called background forces $\widetilde{\chi}_{\text{BG}}$ \cite{borgani2017backgroundSup}. From the measurements of $(\widetilde{d}_{\text{free}},\widetilde{d}_{\text{eng}},\widetilde{d}_{\text{lift}})$ we estimate  the tip-sample nonlinear force at intermodulation frequencies  by applying \cite{hutter2010reconstructingSup}:
\begin{equation}
   \widetilde{F}_{\text{ts}}^{(\text{c})}(\omega)  =  k \bigg [-\frac{\omega}{\omega_0}^2 + j \frac{\omega}{\omega_0}+ 1 \bigg ] \bigg (  \widetilde{d}_{\text{eng}}(\omega) - \widetilde{d}_{\text{free}}(\omega) \bigg ) - \widetilde{\chi}_{\text{BG}}^{-1} \widetilde{d}_{\text{eng}},
\end{equation}in which the last term corresponds to the background force compensation, with its associated linear transfer function defined by \begin{equation}
    \widetilde{\chi}_{\text{BG}}^{-1}(\omega) = k \bigg [-\frac{\omega}{\omega_0}^2 + j \frac{\omega}{\omega_0}+ 1 \bigg ] \bigg ( \frac{\widetilde{d}_{\text{lift}}-\widetilde{d}_{\text{free}}}{\widetilde{d}_{\text{lift}}} \bigg )\label{eq:BGForm}
\end{equation}and approximated on the narrow frequency band with the polynomial \cite{borgani2017backgroundSup}:\begin{equation}
    \widetilde{\chi}_{\text{BG}}^{-1}(\omega) \approx k(a \omega^2 + j b \omega).
\end{equation}The coefficients $a$ and $b$ come from the fit of Eq. \eqref{eq:BGForm} at the two drive frequencies $(\omega_1,\omega_2)$.
In addition, we apply the following phase rotation to compensate the phase shift potentially caused by a time delay inherent to the processing equipment \cite{platz2013reconstructingSup}:\begin{equation}
    \widetilde{F}_{\text{ts,exp}}(\omega)  = \widetilde{F}_{\text{ts}}^{(\text{c})}(\omega)  \text{e}^{-j(R_0 + R_1 \omega/\omega_c)}\label{eq:ShiftPhase}
\end{equation}where $\omega_c=\frac{1}{2}(\omega_1+\omega_2) \approx \omega_0$, and $ \widetilde{F}_{\text{ts}}^{(\text{c})}$ denotes the tip-sample intermodulation components with  the rotation coefficients $(R_0,R_1)$ adjusted such that $\arg(d_{{\text{eng}}}(\omega_1))=\arg({d_{\text{eng}}}(\omega_2))=0$:\begin{gather}
    R_0 = \arg(d_{\text{eng}}(\omega_1))\label{eq:DefR0} -\frac{\arg({d_\text{eng}}(\omega_2))-\arg(d_{\text{eng}}(\omega_1))}{\omega_2-\omega_1}\omega_1 \\
    R_1 = \frac{\arg(d_{\text{eng}}(\omega_2))-\arg(d_{\text{eng}}(\omega_1))}{\omega_2-\omega_1} \omega_c \text{.}\label{eq:DefR1}
\end{gather}The phase equalization procedure  defined by Eqs.~\eqref{eq:ShiftPhase}-\eqref{eq:DefR1} is also applied on the simulated components for comparison purposes.

\newpage
\clearpage
\renewcommand{\thefigure}{S2.\arabic{figure}}
\renewcommand{\thetable}{S2.\arabic{table}}
\section{\label{numeric}Additional numerical data}
\label{sec:S2}
\subsection{Simulations}
\setcounter{figure}{0}

The driving force signal $F_\text{d}(t)$ used in the simulations is defined specifically for the experimental data considered in the study. In particular, it is estimated for the set of frequency, stiffness and quality factor of the first resonance of the cantilever $f_0, k$ and $Q$ obtained from the thermal calibration. The excitation signal is obtained from the free motion frequency components as  :\begin{equation}
    F_{\text{d}}(t) = \sum_{\omega \in \omega_{\text{IM}}} 2 |\widetilde{F}_{\text{d}}(\omega)| \cos(\omega t + \arg(\widetilde{F}_{\text{d}}(\omega)))
    \end{equation}
    with
    \begin{equation}\widetilde{F}_{\text{d}}(\omega) = k \big [-\frac{\omega}{\omega_0}^2 + j \frac{\omega}{Q \omega_0 }+ 1 \big ]  \widetilde{d}_{\text{free}}(\omega)
\end{equation}where the $\omega_{\text{IM}}$ denotes the pulsation of intermodulation \cite{platz2013interpretingSup}.

The time signals are simulated using the following dimensionless values:	\begin{equation} \overline{d}_c  = \frac{d_c}{A} , \quad \overline{d}_s  = \frac{d_s}{A} , \quad \overline{F}_d  = \frac{F_d}{kA}, \quad \overline{F} _{\text{ts}} = \frac{F_{\text{ts}}}{kA}, \quad \overline{t} = \omega_0 t , \quad \overline{h}  = \frac{h}{A}, \quad \overline{s}  = \frac{s}{A}. \label{eq:AdimVal}
\end{equation}
in which the displacement of reference is the amplitude of the engaged motion at the second drive frequency $A=|d_{\text{c}}|_{\omega=\omega_2}$. 
The following dimensionless design parameters are considered in the numerical procedure:\begin{equation} \overline{F}_{\text{ad}}  = \frac{F_{\text{ad}}}{kA}, \quad \overline{k}_v  = \frac{k_v}{k} , \quad \overline{k}_s  = \frac{k_s}{k}, \quad \overline{\eta}_v  = \frac{\eta_v \omega_0}{k} ,  \quad \overline{\eta}_s  = \frac{\eta_s \omega_0}{k}. \label{eq:ParamDimLess}
\end{equation}Thus, the equation of motion (Eq.(1) of the main manuscript) is 
\begin{equation}
\ddot{d}_{\text{c}} + \frac{\dot{d}_{\text{c}}}{Q}  +d_{\text{c}} = F_{\text{d}}(t) + F_{\text{ts}}(s,\dot{s})\label{eq:SimAdim}
\end{equation}in which the overbars are dropped for the sake of brevity.
    The time signals are computed by simulating Eq. \eqref{eq:SimAdim} using a Runge-Kutta scheme. At low sample relaxation times $\overline{\tau_s}=\overline{\eta}_s/\overline{k}_s < 10^{-3}$, a scheme designed for stiff systems is employed (the $ode23s$ function of Matlab is used, instead of the classical $ode45$ time integration solver). 
The signals for $d_c$, $d_s$ and $F_{\text{ts}}$ are simulated on 8 ms, which corresponds to four intermodulation beatings since $\Delta f = \frac{\omega_2 - \omega_1}{2\pi} = 500$ Hz is applied in experiments. A zero initial condition for displacements and velocities is applied.

\begin{figure}[h!]
\begin{scriptsize}
\begin{center}\def\svgwidth{16.8cm}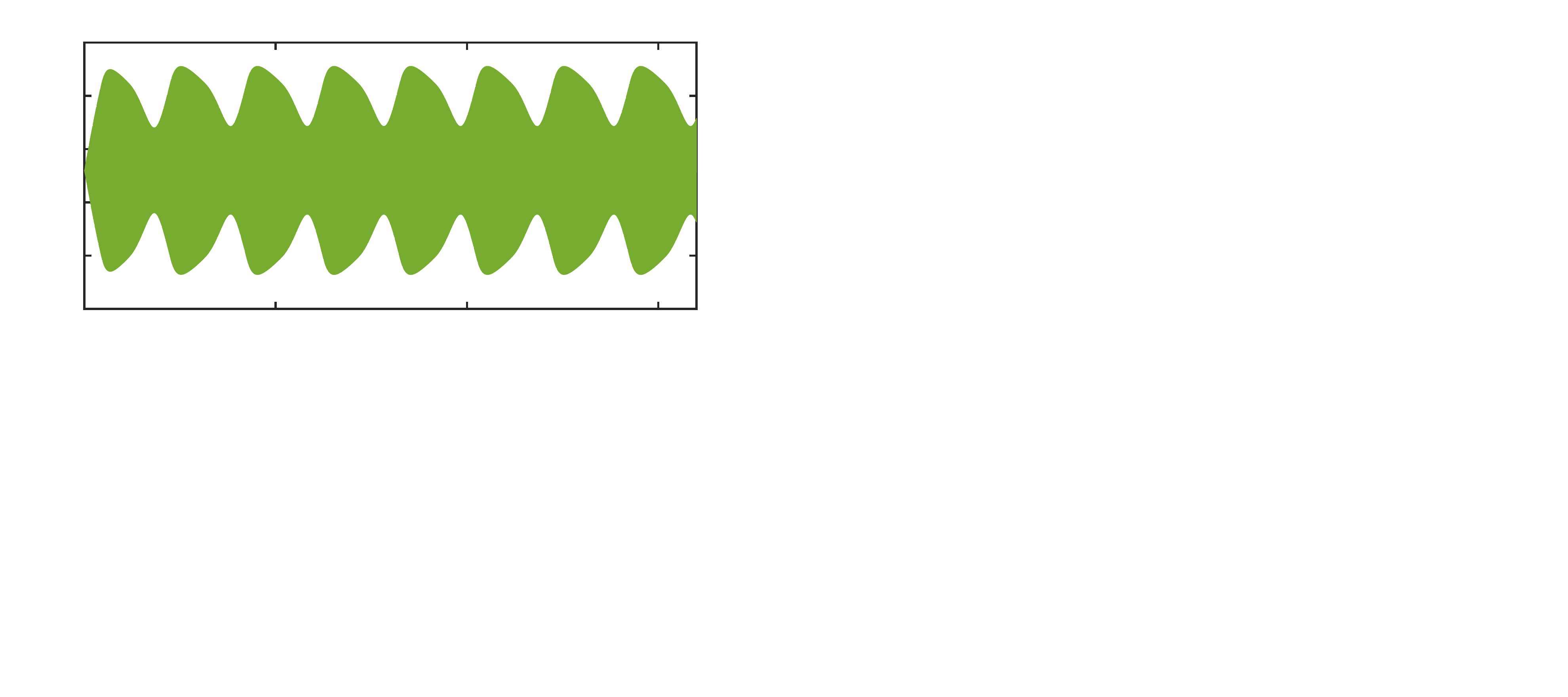
\end{center}\caption{Example of simulated tip displacement (a) and tip-sample force (b) signals. Zoom on the extracted portion of the force signal used for the estimation of the observables (c). Comparison between the direct and averaged amplitudes of the interaction force at the two drive frequencies $\omega_1$ and $\omega_2$ (d).}
\label{fig:FigAnaly}
\end{scriptsize}
\end{figure}
In order to convert the simulated tip-sample force signal $F_{\text{ts,sim}}$ from time to frequency domain at intermodulation frequencies, we extract two beat periods in steady state oscillations (Fig.~\ref{fig:FigAnaly} (a-b)). 
The amplitude and phase components $(|\widetilde{F}_{\text{ts,sim}} |,\phi_{F_{\text{ts}}})$ of 32 intermodulation frequencies are estimated 
using a synchronous detection scheme \cite{tholen2011noteSup}. Next, we use a sliding window with a length equal to one beat period as shown in Fig.~\ref{fig:FigAnaly} (c) and take 10 estimations of the phase and amplitude components. The estimations are then averaged to reduce numerical noise as shown in Fig.~\ref{fig:FigAnaly}(d). 
Finally, the spectral components of the interaction force are stored in the same way as in experiments, in complex form like $\widetilde{F}_{\text{ts,sim}} = |\widetilde{F}_{\text{ts,sim}} |\text{e}^{j \phi_{F_{\text{ts}}} }$. 



The objective function used for estimating the viscoelastic parameters is defined by \cite{forchheimer2012modelSup,platz2013interpretingSup}:
\begin{equation}
f(\bm{P}) = \sqrt{ \sum_{\omega = \omega_{Im}} |\widetilde{F}_{\text{ts,exp}}(\omega) - \widetilde{F}_{\text{ts,sim}}(\omega,\bm{P})|^2 }.
\label{eq:fp}
\end{equation} 
In order to minimize Eq.~\eqref{eq:fp}  we use Levenberg-Marquardt algorithm \cite{levenberg1944methodSup} and combine it with nonlinear least squares (\textit{lsqnonlin} function) in Matlab. This least-square minimization is performed with an iterative procedure  which involves the computation of the partial derivatives (gradient) at each iteration. 
A parallel implementation on a small cluster
was used to perform multiple minimization routines: approximately 10 nodes and 36 hours in total were needed to obtain the results shown in Fig.~5 of the main manuscript. We show in table \ref{Tab:Range} the lower and upper limit of parameter values defined for the optimization. These parameter ranges are deliberately wide because we assume that we have no prior knowledge of the material properties, except in the case of the probe height for which a first approximation can be extracted from the force quadrature curves (see section S2.3).

\setlength\extrarowheight{5pt}
\begin{table}[!ht]
\begin{center}
  \begin{tabular}{| c | c | c | c | c | c | c |}
    \hline
Parameter & $F_{ad}$ [nN] & $k_{v}$ [N.m$^{-1}$]  &  $\eta_{v}$ [mg.s$^{-1}$] & $k_{s}$ [N.m$^{-1}$] &  $\eta_{s}$ [mg.s$^{-1}$] &  $h$ [nm]  
 \\ \hline
 Minimum value &   0.05 	& $\approx 0$   & $\approx 0$& $\approx 0$  & $\approx 0$ & 5 \\ \hline 
 Maximal value &  100 &  $10 k$  & $10 k/\omega_0$ &   $20 k$&  $20k/\omega_0$  & 45 \\\hline 
  \end{tabular}\caption{Parameter ranges used for the optimization routine. Here, $k$ represents the cantilever stiffness in N/m and $\omega_0$ represents the first resonance frequency in rad/s}\label{Tab:Range}
\end{center}
\end{table}

\newpage
\clearpage
\subsection{Results of global optimization tests on synthetic data}

In this section we discuss the use of a global optimization procedure for parameter estimation and further elaborate on the limitations of the procedure. In general global optimization techniques such as Particle swarm optimization  does not rely on gradient descent method used by local optimization techniques like the  Levengerg-Marquardt method, and hence don't require a differentiable  objective function. Such a characteristic helps to determine if the lack of sensitivity of surface motion can be attributed to the chosen optimization algorithm or it is linked to  model parameters. Additionally a global optimization method has the advantage that a large parameter space can be searched from different initial starting points without having prior knowledge on the optimum solution. However, in order to obtain a physically interpretable solution and to reduce the computational time, it is necessary to restrict the search range. We achieve this by assigning values for each of the model parameter from previous experimental characterizations and then extending their ranges by an order of magnitude\cite{Thoren2018Sup,haviland2015probingSup,Penning2020Sup}.

In particular, we choose the sample parameters suitable for PS-LDPE material and generate synthetic data sets based on the interaction with a Silicon cantilever. The sample properties used for the simulations is provided in table \ref{tab:PSO} together with the following cantilever properties: $f_{0}$~=~\SI{163}{\kilo \hertz}, $Q$~=~491, $k$~=~\SI{23.95}{N \per \meter}, the effective driving force $F_{d}$~=~\SI{1.39}{\nano N} and the unperturbed height $h$~=~\SI{22.6}{\nano \meter}. Next, we use random sampling to select different starting parameter sets. A total of 15 different parameter sets are created and simulated with the moving surface model to generate the amplitude and phase frequency components which are then used as inputs for the Particle swarm based global optimization. For all the 15 data sets, the optimization procedure is performed starting from the same initial \virgolette{swarm}. 

\begin{figure}[!ht]
\begin{center}
\includegraphics[width=0.99\columnwidth]{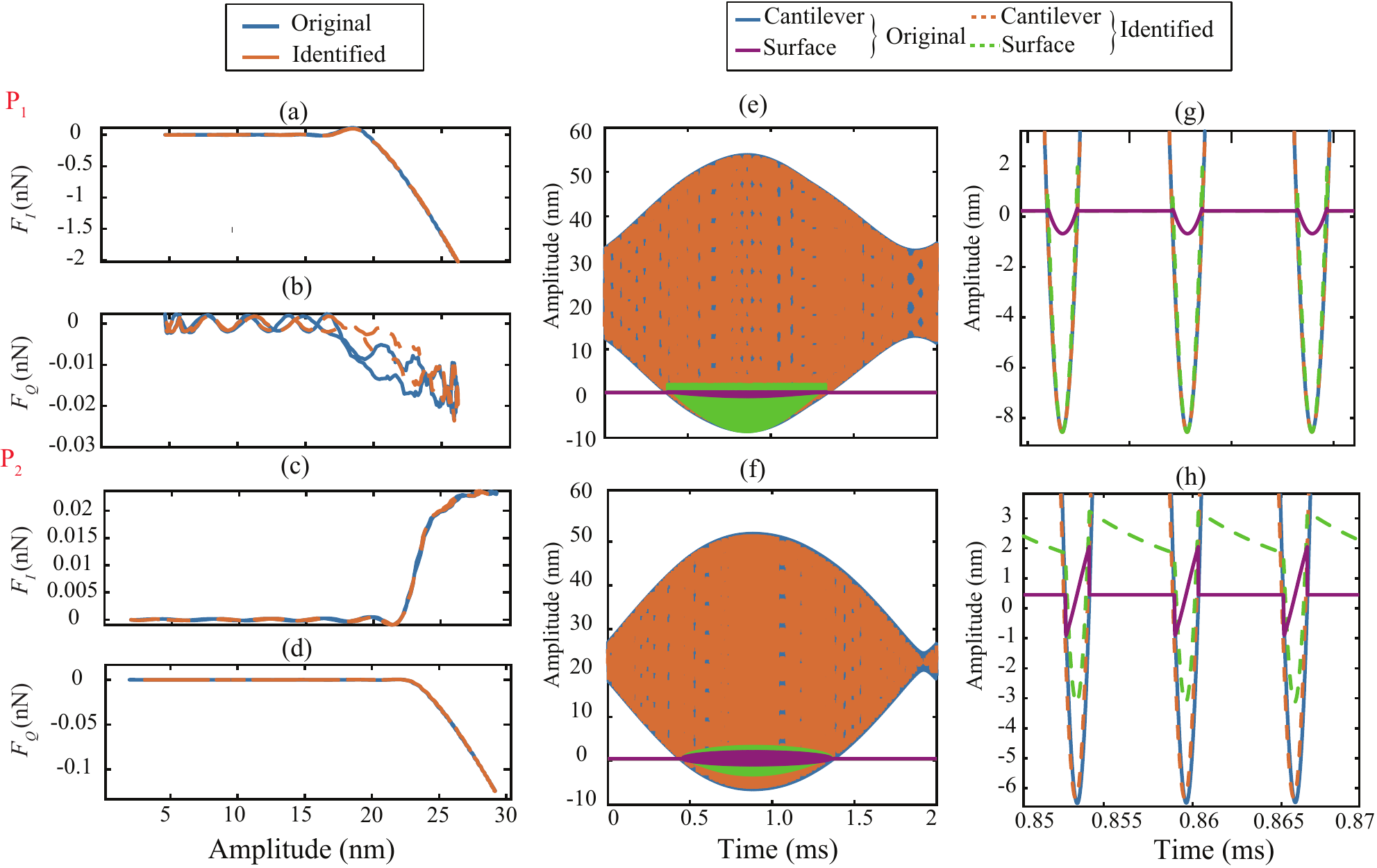}
\end{center}
\caption{Global optimization results for parameter sets 1 and 2. Cantilever properties: $f_{0} = 163$ kHz, $Q = 491$, $k = 23.95$ N/m. Scanning properties: $F_{d} = 1.39$ nN, $h = 22.6$ nm, and 41 amplitude and phase intermodulation products. (a)-(d) Force quadratures showing the conservative and dissipative tip-sample interactions. The blue color represents the original quadratures obtained from model simulations and the orange color represents the identified quadratures based on optimization.  (e)-(f) Time data depicting the motion of the cantilever and the corresponding surface motion due to tip-sample interaction. right: (g)-(h) Zoomed surface motion indicating discrepancies between the original and the identified surface dynamics. The blue and purple color represents the original cantilever and surface dynamics obtained from model simulations; whereas, the orange and green color the original cantilever and surface dynamics based on optimization. }
\label{fig:PSO_beats1}
\end{figure}

\begin{figure}[!ht]
\begin{center}
\includegraphics[width=0.99\columnwidth]{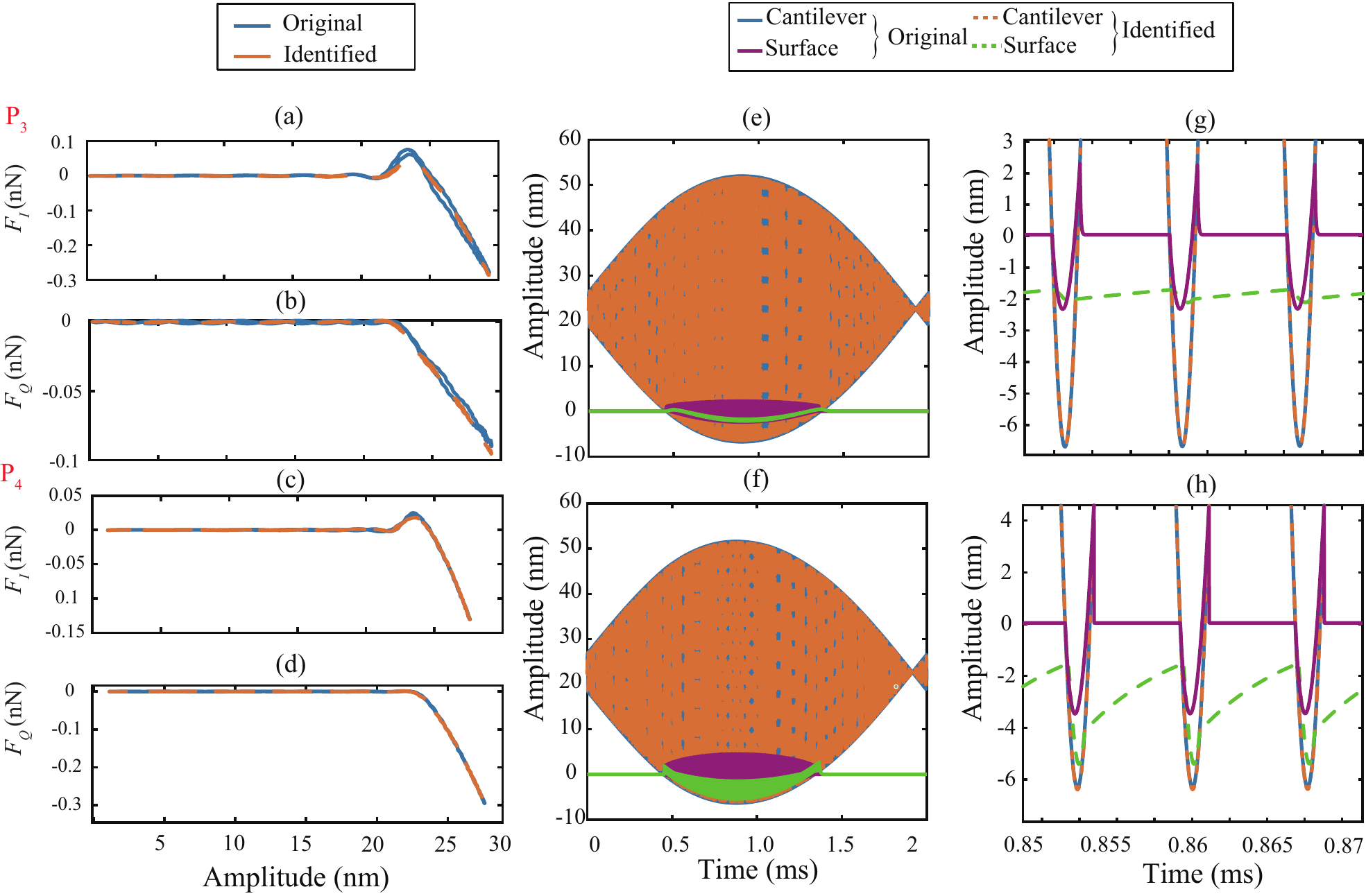}
\end{center}
\caption{Global optimization results for parameter sets 3 and 4. Cantilever properties: $f_{0} = 163$ kHz, $Q = 491$, $k = 23.95$ N/m. Scanning properties: $F_{d} = 1.39$ nN, $h = 22.6$ nm, and 41 amplitude and phase intermodulation products. (a)-(d) Force quadratures showing the conservative and dissipative tip-sample interactions. The blue color represents the original quadratures obtained from model simulations and the orange color represents the identified quadratures based on optimization.  (e)-(f) Time data depicting the motion of the cantilever and the corresponding surface motion due to tip-sample interaction. right: (g)-(h) Zoomed surface motion indicating discrepancies between the original and the identified surface dynamics. The blue and purple color represents the original cantilever and surface dynamics obtained from model simulations; whereas, the orange and green color the original cantilever and surface dynamics based on optimization.}
\label{fig:PSO_beats2}
\end{figure}

Table~\ref{tab:PSO} shows the optimization results for 4 randomly chosen parameter sets out of 15 simulated data sets. The results show that tip-sample dynamics is well approximated with low error values $E$, but the identified parameter values are far from their true values. This deviation is far more significant for surface parameters in comparison with bulk parameters. Once again, we attribute this issue to non-convexity and lack of sensitivity of surface parameters as discussed in the main manuscript. Additionally, Figs.~ \ref{fig:PSO_beats1} and \ref{fig:PSO_beats2} show the temporal data of the cantilever and the associated surface motion together with the force quadratures for both the original dynamics coming from the model simulations and the identified dynamics resulting from optimization. In both the figures, while we observe a good agreement for the force quadratures, the identified motion of the sample surface does not match with the simulated motion (See Figs~\ref{fig:PSO_beats1}(g)-(h) and \ref{fig:PSO_beats2}(g)-(h) ). This further confirms the trivial contribution of the  surface motion on amplitude and phase of intermodulation components.

\begin{quote}
\begin{table}[!ht]
\begin{adjustwidth}{-0.12\textwidth}{-0.12\textwidth}
\begin{center}
  \begin{tabular}{| c | c | c | c | c | c | c | c | c |}
    \hline
Parameter Set & Designation  & $F_{ad}$ [nN] & $k_{v}$ [N.m$^{-1}$]  &  $\eta_{v}$ [mg.s$^{-1}$] & $k_{s}$ [N.m$^{-1}$] &  $\eta_{s}$ [mg.s$^{-1}$] & $E$ (nN)
 \\ \hline
 \multirow{3}{*}{$P_1$} & Optimum & 2.98 & 2.60 & 0.199 & 8.31 & 0.0371 & \multirow{3}{*}{$3.80 \cdot 10^{-3}$} \\ \cline{2-7}
  & PSO result & 2.49 	& 2.04 & 0.181 & 81.0 &  2.14 & \\ \cline{2-7} 
  & Error &  16.4 \% 	& 21.3 \%   & 9.17 \% &  875 \% & 5.67e3 \% &  \\ \hline 
 \multirow{3}{*}{$P_2$} & Optimum & 0.161	& 0.0101   & 0.141 & 0.220 &   1.51  & \multirow{3}{*}{$2.54\cdot 10^{-4}$} \\ \cline{2-7} 
  & PSO result & 0.165	& 0.0100 & 0.135 & 16.8 &   0.00155  & \\ \cline{2-7} 
  & Error &  3.00 \% 	& 0.547 \%   & 4.05 \% &  7.53e3 \% & 99.9 \% &  \\ \hline
 \multirow{3}{*}{$P_3$} & Optimum & 4.49 	& 6.81 & 0.0221 & 0.108 & 0.582 & \multirow{3}{*}{$2.11 \cdot 10^{-2}$} \\ \cline{2-7} 
  & PSO result &  8.18 	& 0.97   & 0.378 &  0.938 & 0.00105 & \\ \cline{2-7} 
  & Error &  39.6 \% 	& 25.7 \%   & 20.0 \% &  680 \% & 99.0\% &  \\ \hline 
 \multirow{3}{*}{$P_4$} & Optimum & 0.473	& 0.349 & 0.469 & 65.5 & 0.0105 & \multirow{3}{*}{$8.12\cdot 10^{-4}$} \\ \cline{2-7} 
  & PSO result & 0.277	& 0.283 & 0.802 & 1.20 &  0.0360  & \\  \cline{2-7} 
  & Error &  41.5 \% 	& 19.0 \%   & 71.1 \% &  98.2 \% & 245 \% &  \\ \hline  
  \end{tabular}
\end{center}
\end{adjustwidth}
\caption{Parameter Convergence for data sets P4, P7, P9 and P13. Cantilever properties used: $f_{0} = 163$ kHz, $Q = 491$, $k = 23.95$ N/m. Scanning properties: $F_{d} = 1.39$ nN, $h = 22.6$ nm, and 41 amplitude and phase intermodulation products.}\label{tab:PSO}
\end{table}
\end{quote}

\newpage
\clearpage
\subsection{Criterion for probe height identification from force quadratures}

\begin{figure}[!ht]
\begin{center}
\includegraphics[width=0.8\columnwidth]{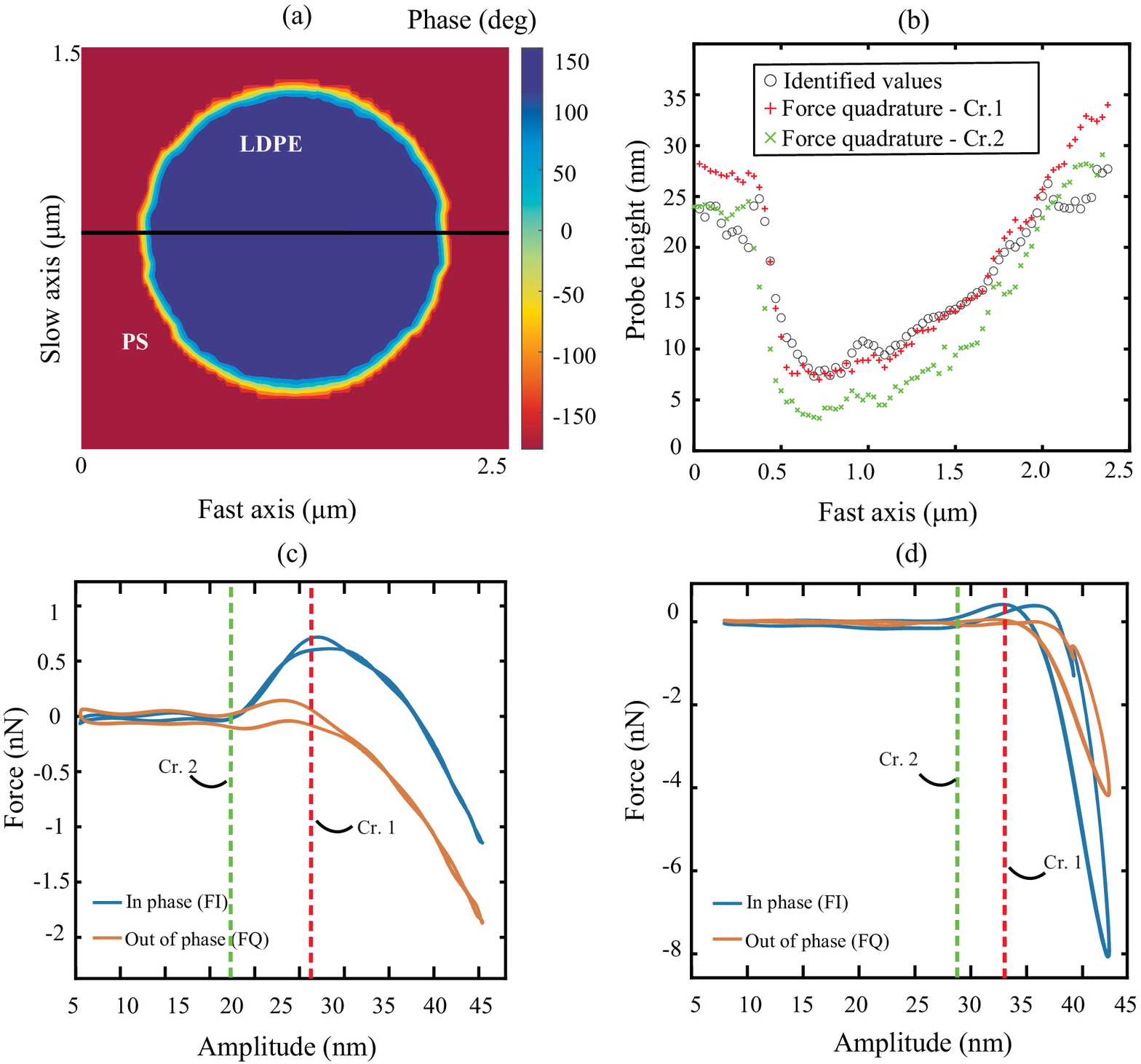}
\end{center}
\caption{Portion of the extracted line for the analysis in the AFM image (top left). Identified $h$ and reported values directly read from the force quadratures (top right). Illustration of the two criteria (dashed lines) for estimating the probe height on the quadratures on one pixel made of PS (bottom left) and of LDPE (bottom right).}
\label{fig:himageAndCurve}
\end{figure}

The probe height $h$ is included in the set of unknown parameters (see main manuscript modelling section 3). In general \textit{h} varies with the working height of the cantilever which in turn depends on how much the feedback control moves the z-piezo during the scanning operation. By taking advantage of the conservative quadrature,  in phase with the cantilever motion, it is possible to estimate an approximate value for \textit{h} based on the onset of repulsive forces.

We suggest two criteria for extracting \textit{h} from force quadratures as illustrated in Fig.~\ref{fig:himageAndCurve}.
We assume the maximum of the in-phase force component (related to adhesion) is achieved closely after the tip starts to penetrate the sample. Thus, the first criterion (denoted by red crosses in Fig. \ref{fig:himageAndCurve} (b)) is taken at the middle of the increasing part of $F_I$, whereas the second one corresponds to the amplitude where the in-phase component starts to increase. 
We browse and apply these two criterion on all pixels of the black line displayed in Fig.~\ref{fig:himageAndCurve}(a). The comparison shown in Fig.~\ref{fig:himageAndCurve}(b) highlights a better match between the heights corresponding to LDPE pixels using the first criteria, when the second criteria seems more suited for the pixels related to PS material. That can be explained by the different material properties, for instance the larger stiffness for PS causes a faster increase of $F_I$, whereas in case of the softer material the short-range adhesive force is more significantly involved before the tip starts to indent the sample. The analysis of these force quadrature curves could be further developed using a more accurate tip-sample force model such as Attard's model \cite{attard2001interactionSup,attard2007measurementSup,rajabifar2018dynamicSup,rajabifar2021discriminationSup,rajabifar2021fastSup}, in order to describe first the transition between the non-contact and adhesive regime, and secondly the transition between the adhesive and repulsive regime. 

\newpage
\clearpage
\renewcommand{\thefigure}{S3.\arabic{figure}}
\renewcommand{\thetable}{S3.\arabic{table}}

\section{High volume gradient based optimization and initial point selection procedure\label{sec:IPChoice}}

In this section, we discuss the results obtained using the Levenberg-Marquardt algorithm from multiple initial points for both models with and without sample's surface motion. This is done to analyze the sensitivity of the model on initial starting points for the optimization. We begin by creating a numerical range for each parameter based on previous literature studies. Then, a grid of initial starting points is chosen and for each initial point we perform the optimization routine. The distribution of the identified parameters is analysed with histograms and  by fitting Gaussian function to extract statistics. The distribution  are discussed for each  model separately in the following sections. 

\subsection{\label{sec:simpara}Piecewise linear model with surface motion}
Using the moving surface model, we run multiple gradient-based optimizations for pixel (i) and pixel (iii) of Fig.~2 in the main manuscript with the  grid of initial parameters defined in table \ref{Tab:PWL_IP}. The grid includes 3 different values per parameters, chosen in such a way that  the parameter exploration recovers a large parameter space (including notably at least one order of magnitude in the case of the viscoelastic properties), and that all routines are performed within a reasonable computational time.  In total, $3^6=729$ optimizations were performed, starting from all the combinations of the grid.
In this section  we present the histograms used to extract the values reported in Table~1 of the main manuscript.
\begin{table}[h!]
\begin{center}
  \begin{tabular}{| c | c | c | c | c | c |}
    \hline
 $F_{ad}$ [nN] & $k_{v}$ [N.m$^{-1}$] &  $\eta_{v}$ [mg.s$^{-1}$] & $k_{s}$ [N.m$^{-1}$] &  $\eta_{s}$ [mg.s$^{-1}$] &  $h$ [nm]  \\ \hline
    [5 25 45]	& [0.02 1 40]  & [0.2 1 5] & [0.02 1 40]  & [0.2 1 5] & [15 25 35] \\ \hline
  \end{tabular}\caption{Grid of initial points for the local optimization procedure using the moving surface model.}\label{Tab:PWL_IP}
\end{center}
\end{table}

\begin{figure}[h!]
\begin{center}
\includegraphics[width=1.0\columnwidth]{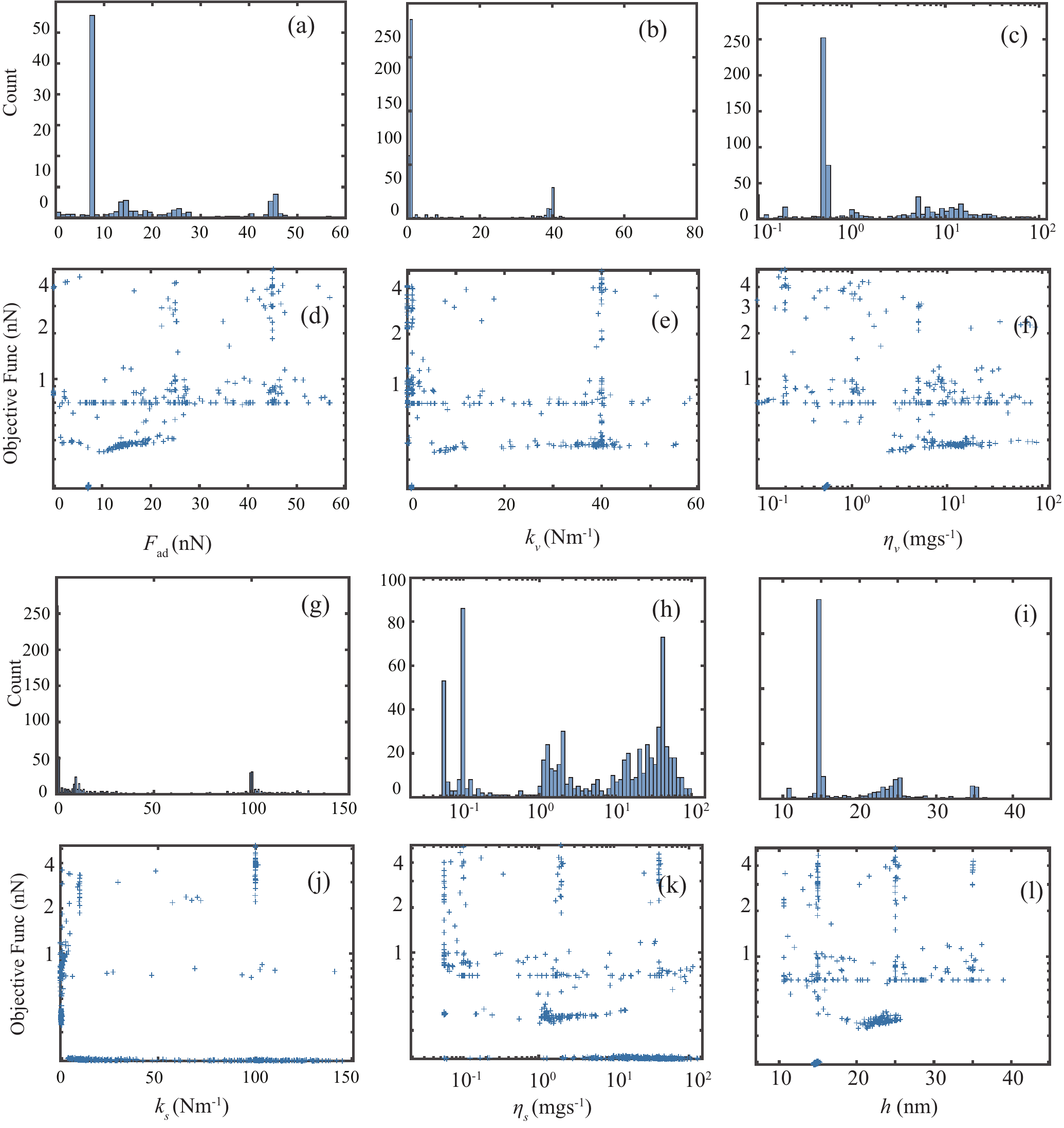}
\end{center}
\caption{Identified parameters of the PWL model with sample motion, obtained on LDPE material at pixel (iii) of Fig. 2(b) in the main manuscript with the initial positions defined in table \ref{Tab:PWL_IP}. Parameter distributions and errors are respectively plotted in (a)\&(d) for $F_a$, (b)\&(e) for $k_v$, (c)\&(f) for $\eta_v$, (g)\&(j) for $k_s$, (h)\&(k) for $\eta_v$ and (i)\&(l) for $h$.}\label{fig:DiscussIP_PWL5Par_iii}
\end{figure}

\begin{figure}[h!]
\begin{center}
\includegraphics[width=1.0\columnwidth]{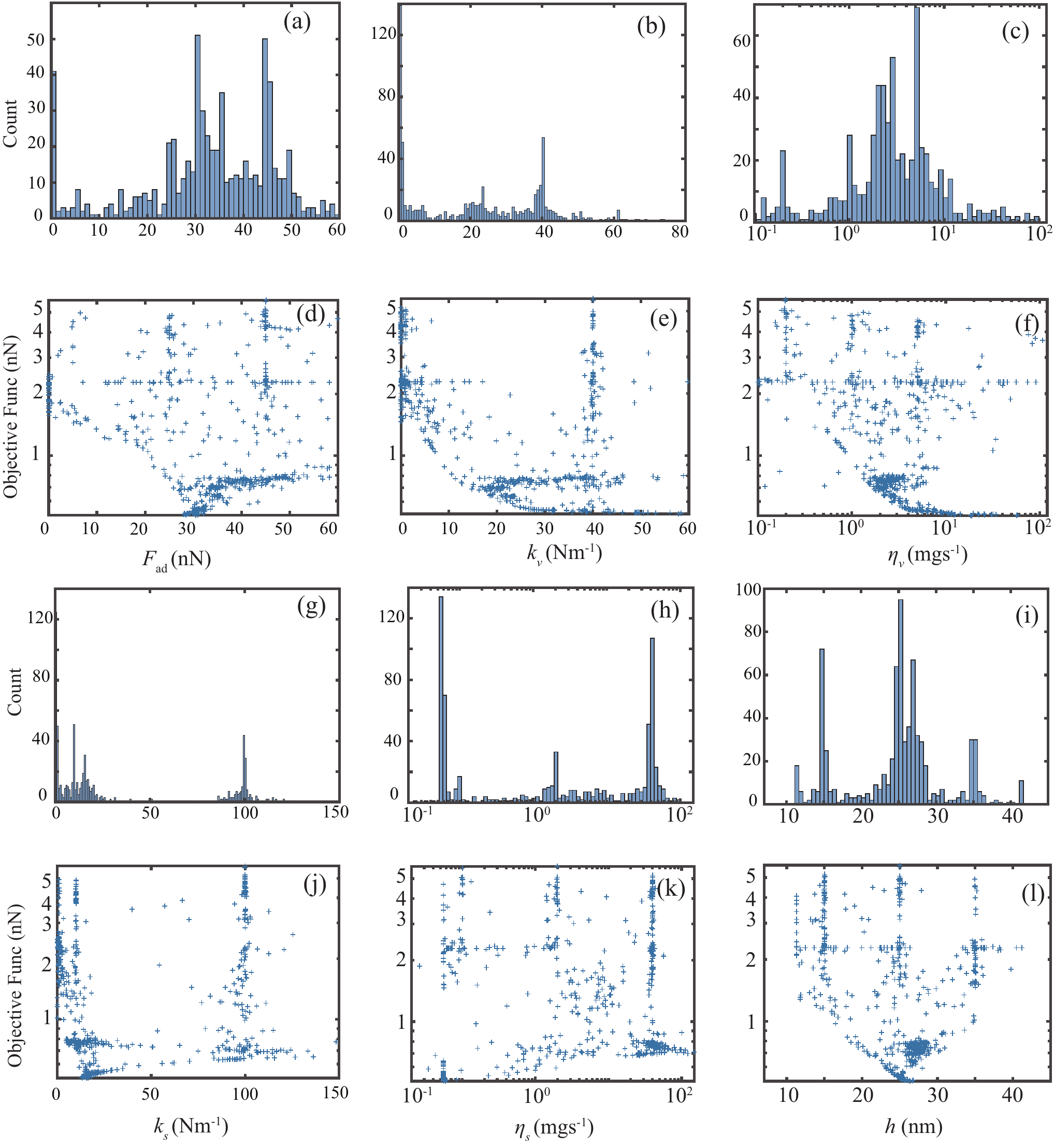}
\end{center}
\caption{Identified parameters of the PWL model with sample motion, obtained on PS material at pixel (i) of Fig. 2(b) in the main manuscript with the initial positions defined in table \ref{Tab:PWL_IP}. Parameter distributions and errors are respectively plotted in (a)\&(d) for $F_a$, (b)\&(e) for $k_v$, (c)\&(f) for $\eta_v$, (g)\&(j) for $k_s$, (h)\&(k) for $\eta_v$ and (i)\&(l) for $h$.}\label{fig:DiscussIP_PWL5Par_i}
\end{figure}

Figures~\ref{fig:DiscussIP_PWL5Par_iii} and \ref{fig:DiscussIP_PWL5Par_i} highlight the distribution of the identified parameters with respect to the objective function for pixels (i) and (iii), respectively. We see a clear correlation between a large distribution and low errors only for some parameters such as $F_a$, $k_v$, $\eta_v$, $h$ for pixel (iii) in Fig.~\ref{fig:DiscussIP_PWL5Par_iii}. If   model parameters have strong correlation with  the objective function then the maximum of the histogram counts (rows 1 and 3) coincides with the minima of the scatter plots (rows 2 and 4). For example, in case of Figs.~\ref{fig:DiscussIP_PWL5Par_iii}(a) and (d), we look at the influence of adhesion force $F_a$ on the objective function and we observe that the location of the maximum along the x-axis in Fig.~\ref{fig:DiscussIP_PWL5Par_iii}(a) coincides with the minima along the same x-axis in Fig.~\ref{fig:DiscussIP_PWL5Par_iii}(d). A similar behaviour is observed in   Figs.~\ref{fig:DiscussIP_PWL5Par_iii} (b)\&(e), (c)\&(f), and(i)\&(l). On the contrary, Figs.~\ref{fig:DiscussIP_PWL5Par_iii} (g)\&(j) and (h)\&(k) lack such property and instead exhibit  random and spread distributions. This behavior is due to the insensitivity of the objective function to the sample parameters. A similar observation holds  for the PS material (Fig.~\ref{fig:DiscussIP_PWL5Par_i})    with an even more complex  distribution. It is here attributed to the combined effect of non-convexity and insensitive regions in which the optimizer encounters a stopping condition.

\newpage
\clearpage
\subsection{Piecewise linear model without surface motion}\label{subsubsec:Stat_4Params}

Here, we report the results and histograms obtained from the large set of optimizations carried out using the piecewise linear model without surface motion. We begin with a set of $3^4$ initial parameters defined by the grid presented in table \ref{Tab:grid_MSM_wo_MS}, and analyze the parameter distributions in the same way as outlined in the previous section.

With the 4 parameters model,  statistic  for the identified parameters   depicts   well defined Gaussian distributions that are specific for each type of material. Additionally, the mean of the Gaussian distributions correspond to the lowest values of the objective function. This is shown in Figs.~\ref{fig:DiscussIP_PWL5Par_iii} and \ref{fig:DiscussIP_PWL5Par_i} for   PS and LDPE material sampled at pixel locations (i) and (iii) of Fig.~2 in the main manuscript. The parameter values from the optimization procedure  are reported in table~\ref{Tab:MeanUncert_IP_PWL3}.

\begin{table}[h!]
\begin{center}
  \begin{tabular}{| c | c | c | c |}
    \hline
 $F_{ad}$ [nN] & $k_{v}$ [N.m$^{-1}$] &  $\eta_{v}$ [mg.s$^{-1}$] &  $h$ [nm]  \\ \hline
    [5 25 45]	& [0.02 1 40]  & [0.2 1 5] & [15 25 35] \\ \hline 
  \end{tabular}\caption{Grid of initial points for the local optimization procedure using PWL model without sample motion.}\label{Tab:grid_MSM_wo_MS}
\end{center}
\end{table}

\begin{figure}[h!]
\begin{center}
\includegraphics[width=1.0\columnwidth]{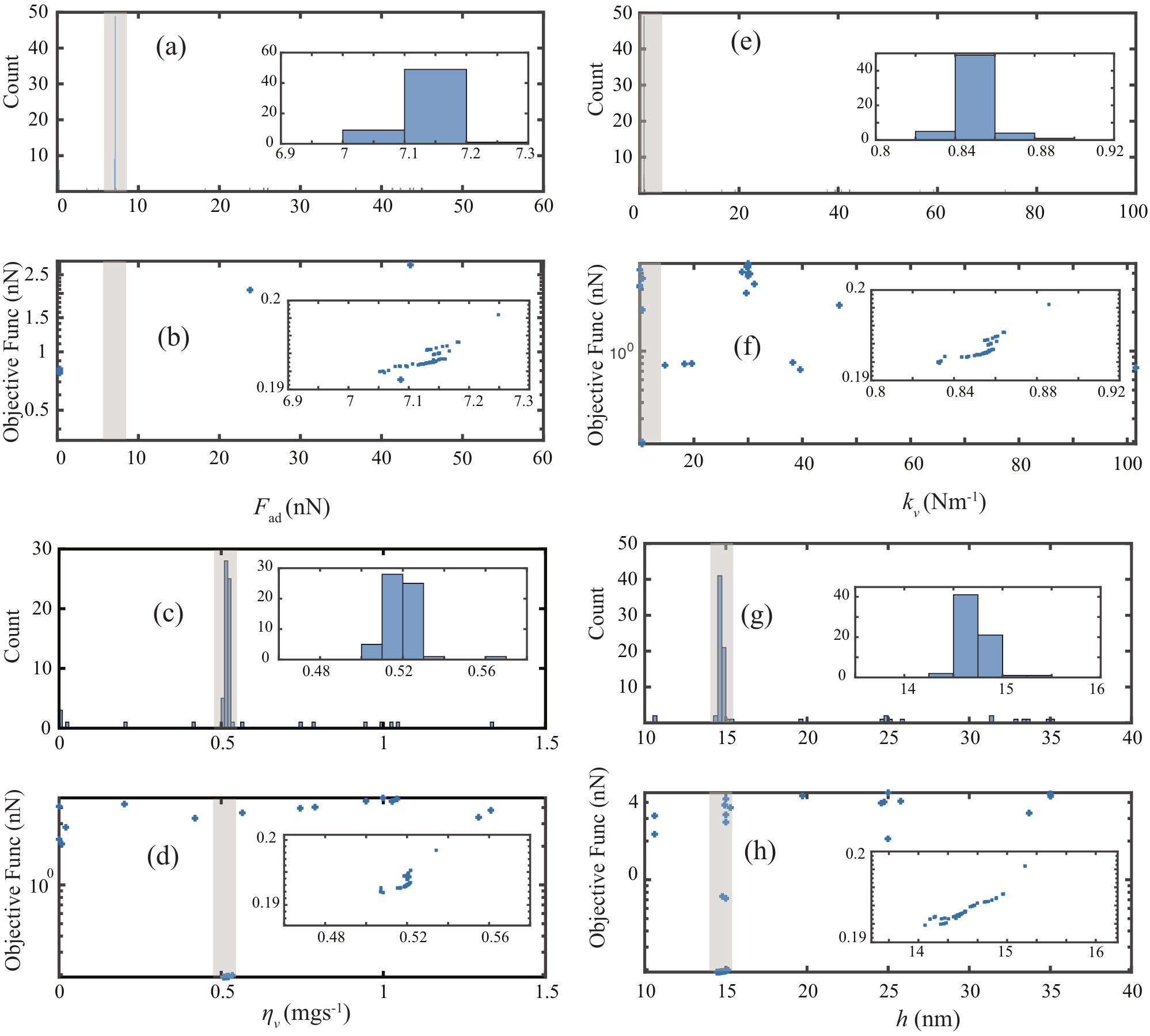}
\end{center}
\caption{Identified parameters of the PWL model without sample motion, obtained on pixel (iii) of Fig. 2(b) in the main manuscript (LDPE) from the initial positions defined in table \ref{Tab:grid_MSM_wo_MS}. Parameter distributions and errors are respectively plotted in (a)\&(b) for $F_a$, (e)\&(f) for $k_v$, (c)\&(d) for $\eta_v$ and (g)\&(h) for $h$. The shadowed areas highlight the Gaussian distributions.
}
\label{fig:DiscussIP_PWL3Par_iii}
\end{figure}

\begin{figure}[h!]
\begin{center}
\includegraphics[width=1.0\columnwidth]{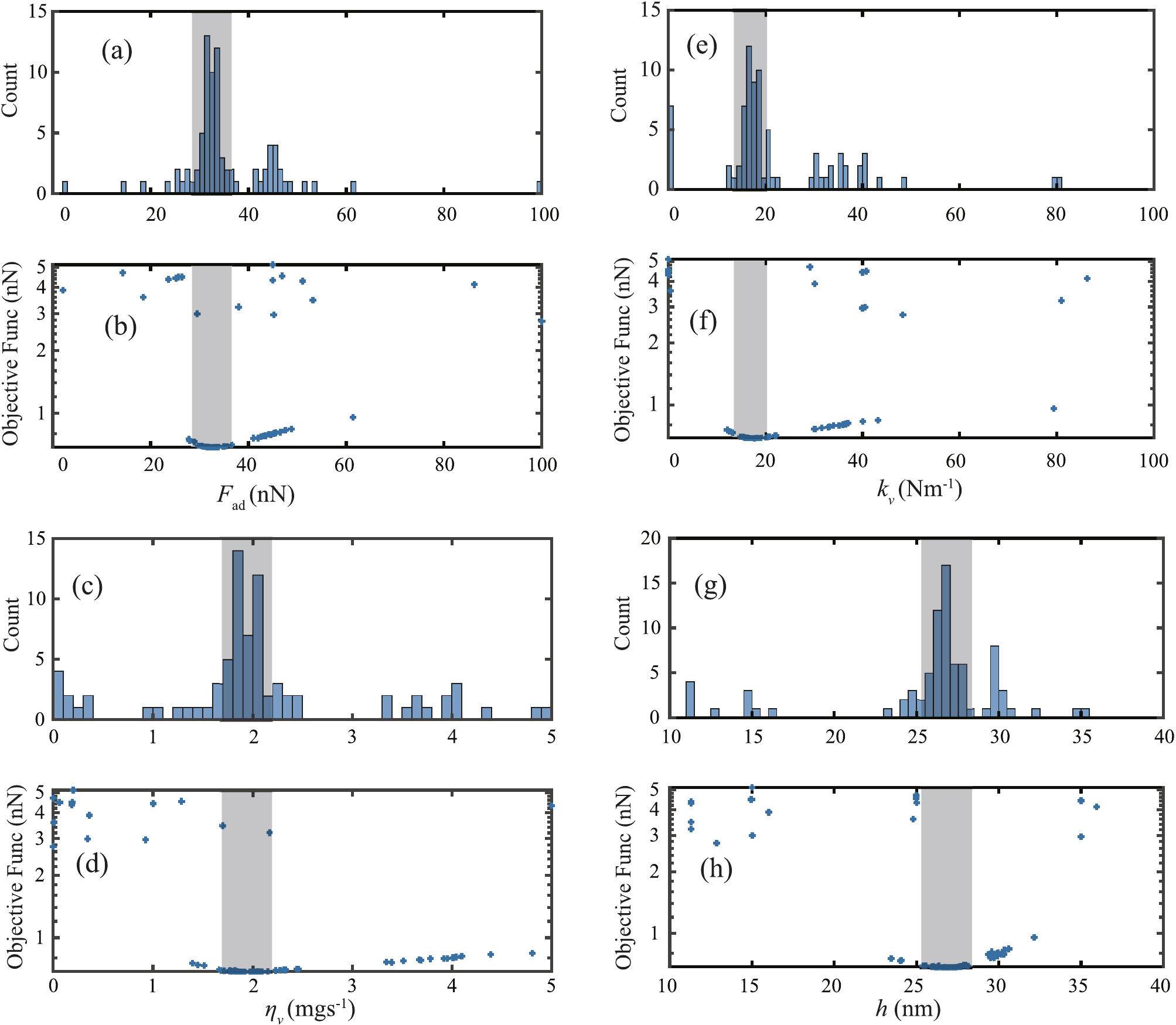}
\end{center}
\caption{Identified parameters of the PWL model without sample motion, obtained on pixel (i) of Fig. 2 in the main manuscript (PS), starting from the initial positions defined in table \ref{Tab:grid_MSM_wo_MS}. Parameter distributions and errors are respectively plotted in (a)\&(b) for $F_a$, (e)\&(f) for $k_v$, (c)\&(d) for $\eta_v$ and (g)\&(h) for $h$. The shadowed areas highlight the Gaussian distributions.
}
\label{fig:DiscussIP_PWL3Par_i}
\end{figure}

\begin{table}[ht!]
\centering
  \begin{tabular}{| c | c | c |}
  \hline
 & Pixel (i) & Pixel (iii) \\ \hline 
$F_{ad}$ [nN] & $ 32.7 \pm  0.45 $ & $ 7.13\pm 0.008 $ \\ \hline
 $k_{v}$ [N/m]  & $17.52 \pm  0.52$ & $0.854\pm  0.002$ \\ \hline
$\eta_{v}$ [mg/s]  & $1.975\pm  0.006$ & $0.519 \pm  0.001$ \\ \hline
$h$ [nm]  & $26.7 \pm 0.18 $ & $14.7 \pm  0.03 $    \\ \hline
  \end{tabular}
    \caption{Identified parameters resulting from the Gaussian fits. We extracted the results with errors smaller than 0.71 nN for pixel (i) (cf Fig.~\ref{fig:DiscussIP_PWL3Par_i}) and 0.25 nN for pixel (iii) (cf Fig.~\ref{fig:DiscussIP_PWL3Par_iii}). The uncertainties are estimated with a 95\% confidence interval.
    }\label{Tab:MeanUncert_IP_PWL3}
\end{table}

From this statistical analysis, we
extract a reduced set of starting parameters. The parameters summarised  in table \ref{Tab:IPsOnly_PWL4}  have been used  to obtain the results showcased in Fig.~5 of the main manuscript.
The two first initial points in Table \ref{Tab:IPsOnly_PWL4} were selected by identifying the mean values (also corresponding with the lowest error) among the final results 
displayed in Figs.~\ref{fig:DiscussIP_PWL3Par_iii} and \ref{fig:DiscussIP_PWL3Par_i}. 
In addition, we add a third initial point leading to identified parameters within the confidence intervals for all parameters and both the pixels.
We detail the final parameters and errors obtained on pixels (i) and (iii) with these three initial points in table~\ref{Tab:IPs_PWL4}.
\begin{table}[htbp!]
\centering
  \begin{tabular}{| c | c | c | c |}\hline
  $F_{ad}$  &  $k_{v} $  & $\eta_{v}$  & $h$  \\ 
  \hline
 [nN] & [N/m] & [mg/s] & [nm] \\ \hline 
 $45$ & 0.02 & 0.2 & 35  \\ 
 $5$ &  1 & 1 & 35  \\ 
 5 & 1 & 1 & 15  \\
 \hline
  \end{tabular}
    \caption{Initial starting parameters used as inputs for the optimization performed on the AFM scan.
    }\label{Tab:IPsOnly_PWL4}
\end{table}

\begin{table}[htbp!]
\centering
\begin{adjustwidth}{-0.12\textwidth}{-0.12\textwidth}
  \begin{tabular}{| c | c | c | c | c | c | c | c | c | c | c | c | c | c |}
  \hline
 \multicolumn{4}{|c|}{}& \multicolumn{5}{|c|}{Pixel (i)} & \multicolumn{5}{|c|}{Pixel (iii)} \\ \hline
 \multicolumn{4}{|c|}{Initial parameters} & \multicolumn{4}{|c|}{Final parameters} & \hspace{-0.2cm} Final Error \hspace{-0.2cm}  & \multicolumn{4}{|c|}{Final parameters} & \hspace{-0.2cm} Final Error \hspace{-0.2cm} \\
  \hline
 \hspace{-0.1cm} $F_{ad}$ \hspace{-0.1cm} &  $k_{v} $  & $\eta_{v}$  & $h$ & $F_{ad}$  &  $k_{v} $  & $\eta_{v}$ & $h$  & $E$& $F_{ad}$  &  $k_{v} $  & $\eta_{v}$ & $h$  & $E$ \\ 
  \hline
\hspace{-0.2cm} [nN] \hspace{-0.2cm} & \hspace{-0.2cm} [N/m] \hspace{-0.2cm} & \hspace{-0.2cm} [mg/s] \hspace{-0.2cm} & [nm] \hspace{-0.2cm}& \hspace{-0.2cm} [nN]\hspace{-0.2cm} & \hspace{-0.2cm} [N/m] \hspace{-0.2cm} & \hspace{-0.2cm} [mg/s] \hspace{-0.2cm} & [nm] \hspace{-0.2cm} & \hspace{-0.2cm}[nN] \hspace{-0.2cm}&  [nN] \hspace{-0.2cm}& \hspace{-0.2cm}[N/m]\hspace{-0.2cm} & \hspace{-0.2cm}[mg/s]\hspace{-0.2cm} & \hspace{-0.2cm}[nm]\hspace{-0.2cm} & [nN]\hspace{-0.2cm} \\ \hline 
 $45$ & 0.02 & 0.2 & 35  & 32.9 & 17.74 & 1.99 & 26.9 & 0.68 & 0.3 & 9.33 & 21.1 & 31.5 & 0.78 \\ 
 $5$ &  1 & 1 & 35 & 33.8 & 18.6 & 2.08 & 27.0 & 0.69 & 7.06 & 0.833 & 0.508 & 14.4 & 0.192 \\ 
 5 & 1 & 1 & 15 & 32.0  & 16.7 & 1.89 & 26.6 & 0.69 & 7.12 & 0.852 & 0.519 & 14.6 & 0.193  \\
 \hline
  \end{tabular}
  \end{adjustwidth}
    \caption{Identified parameters and final errors obtained at pixels (i) and (iii) from the three selected initial points defined in Table~\ref{Tab:IPsOnly_PWL4}.
    }\label{Tab:IPs_PWL4}
\end{table}

\clearpage
\newpage



\end{document}